\documentclass[11pt]{article}
\usepackage[english]{babel}
\usepackage[plain]{fullpage}
\usepackage{float}
\usepackage{lineno} 
\usepackage{graphicx,multicol}
\usepackage{epic,eepic,epsfig}
\usepackage{amssymb}
\usepackage{amsmath,amsthm}
\usepackage{mathrsfs}
\usepackage{color}
\usepackage{tikz, tkz-graph, tkz-berge}
\usepackage{algorithm}
\usepackage{algorithmicx}
\usepackage{algpseudocode}
\usepackage[inline]{enumitem}
 
\renewcommand{\epsilon}{\varepsilon}
\newcommand{\defproblem}[3]{
 \vspace{3mm}
\noindent\fbox{
 \begin{minipage}{0.96\textwidth}
 \begin{tabular*}{\textwidth}{@{\extracolsep{\fill}}lr} #1  \\ \end{tabular*}
 {\bf{Input:}} #2 \\
 {\bf{Output:}} #3
 \end{minipage}
 }
 \vspace{3mm}
}
\newcommand{\varsymb}{\mathbf{x}}

\usetikzlibrary{decorations.pathreplacing}
\usetikzlibrary{patterns}
\usetikzlibrary{patterns.meta}

\definecolor{g-green}{rgb}{0.235, 0.659, 0.322}
\definecolor{g-blue}{rgb}{0.0, 0.5, 1.0}

\newboolean{colouredfigures}
\setboolean{colouredfigures}{true}

\usepackage{eso-pic} 
\theoremstyle{plain}
\newtheorem{theorem}{Theorem}
\newtheorem{conjecture}{Conjecture}
\newtheorem{claim}{Claim}[theorem]
\newtheorem{corollary}[theorem]{Corollary}
\newtheorem{lemma}[theorem]{Lemma}

\theoremstyle{definition}
\newtheorem{remark}[theorem]{Remark}
\newtheorem{problem}[theorem]{Problem}
\newtheorem{question}[theorem]{Question}

\newenvironment{subproof}{\par\noindent {\it Proof}.\ }{\hfill$\lozenge$\par\vspace{11pt}}

\begin{document}
\bibliographystyle{plain}

\title{Constrained Flows in Networks}
\author{J. Bang-Jensen\thanks{Department of Mathematics and Computer
    Science, University of Southern Denmark, Odense, Denmark (email:
    jbj@imada.sdu.dk). Research supported by the Independent
    Research Fund Denmark under grant number DFF 7014-00037B}\and S. Bessy\thanks{LIRMM, Univ Montpellier, CNRS, Montpellier, France (email:stephane.bessy@lirmm.fr), financial supports: PICS CNRS DISCO and ANR DIGRAPHS n.194718.}
    \and L. Picasarri-Arrieta\thanks{Universit\'e C\^ote d'Azur, CNRS, Inria, I3S, Sophia-Antipolis, France (email:lucas.picasarri-arrieta@inria.fr), financial supports: DIGRAPHS ANR-19-CE48-0013 and EUR DS4H Investments ANR-17-EURE-0004.}
    }

\date{}

\maketitle

\sloppy

\begin{abstract}
  The {\bf support} of a flow $x$ in a network is the subdigraph induced by the arcs $uv$ for which $x(uv)>0$.
  We discuss a number of results on flows in networks where we put certain restrictions on structure of the support of the flow. Many of these problems are NP-hard because they generalize linkage problems for digraphs. For example deciding whether a network ${\cal N}=(D,s,t,c)$ has a maximum flow $x$ such that the maximum out-degree of the support $D_x$ of $x$ is at most 2 is NP-complete as it contains the 2-linkage problem as a very special case.
  
  Another problem which is NP-complete for the same reason is that of computing the maximum flow we can send from $s$ to $t$ along $p$  paths (called a maximum  {\bf $p$-path-flow}) in ${\cal N}$. Baier et al. (2005) gave a polynomial time algorithm which finds a $p$-path-flow $x$ whose value is at least $\frac{2}{3}$ of the value of a optimum $p$-path-flow when $p\in \{2,3\}$, and at least $\frac{1}{2}$ when $p\geq 4$. When $p=2$, they show that this is best possible unless P=NP. 
  We show for each $p\geq 2$ that the value of a maximum $p$-path-flow cannot be approximated by any ratio larger than $\frac{9}{11}$, unless P=NP. We also consider a variant of the problem where the $p$ paths must be disjoint. For this problem, we give an algorithm which gets within a factor $\frac{1}{H(p)}$ of the optimum solution, where $H(p)$ is the $p$'th harmonic number ($H(p) \sim \ln(p)$). We show that in the case where the network is acyclic, we can find such a maximum $p$-path-flow in polynomial time for every $p$. 
  
  We determine the complexity of a number of related problems concerning the structure of flows. For the special case of acyclic digraphs, some of the results we obtain are in some sense best possible.
  
  \noindent{}{\bf Keywords:} flows, (arc-)disjoint paths with prescribed end vertices, acyclic digraph, polynomial time algorithm, NP-complete problem, approximation algorithm, parameterised complexity.
  
\end{abstract}

\section{Introduction}

 Flows in networks form a very useful tool for modelling and solving many practical optimization problems. The maximum $(s,t)$-flow problem (where we seek a maximum flow from a source $s$ to a sink $t$) is very well studied and many efficient algorithms are known (for an extensive collection of results on flows, see~\cite{ahuja1993}). This combined with the Max Flow Min Cut theorem which characterizes the value of a maximum $(s,t)$-flow has led to efficient algorithms for numerous optimization problems. Maximum flow algorithms are also often used as subroutines in algorithms for more complicated - not necessarily - polynomial time algorithms for solving complex optimization problems. Another very nice and useful property of flows is that any flow, no matter how we choose the capacities (including choosing irrational numbers) can be decomposed into a linear number of much simpler flows, called {\bf path-flows} and {\bf cycle-flows} (see e.g. Chapter 4 in~\cite{bang2009}).

There are, however, many problems concerning  flows which cannot be solved using standard flow algorithms. The purpose of this paper is to study the complexity of a number of such problems all of which have a natural interpretation.  All flows in this paper will be  integer valued, that is, the flow on every arc is a non-negative integer. One  problem that we study is the so-called maximum {\bf $p$-splittable} flow problem~\cite{baierA42} where the goal is to find the maximum value of an $(s,t)$-flow $x$ which can be decomposed into $p$, not necessarily arc-disjoint, path-flows. 
When $p=2$, Baier et al.~\cite{baierA42} showed that problem is already NP-hard and cannot be approximated better than a factor of $\frac{2}{3}$, unless P=NP.
They also gave a $\rho$-approximation algorithm for the $p$-splittable flow problem, with flows valued in $\mathbb{R}$, where $\rho = \frac{2}{3}$ when $p\leq 3$ and $\rho = \frac{1}{2}$ when $p\geq 4$. As we explain later, using a rounding strategy, this algorithm can be adapted into a $\frac{\rho}{2}$-approximation for the problem with flows valued in $\mathbb{N}$.
In this paper we give a $\frac{1}{H(p)}$-approximation algorithm for this problem, where $H(p)$ is the $p$'th harmonic number\footnote{The $p$'th harmonic number is defined as $H(p) = \sum_{i=1}^p \frac{1}{i}$.}. This is an improvement on the $\frac{\rho}{2}$-approximation above when $p\leq 30$.
We also give non-approximability results for every fixed $p\geq 2$.

A further restriction on $p$-splittable flows could be that the $p$-paths must be arc-disjoint or even vertex-disjoint (apart from the end vertices $s,t$). We show that our approximation algorithm above applies to both of these cases also. 

The paper is organized as follows. In Section~\ref{sec:prelim} we recall some basic properties of flows as well as some classical complexity results. In Section~\ref{sec:degreeconstrained} we study the complexity of flow problems where we want to restrict the maximum number of arcs that carry flow and are all leaving the same vertex. This problem turns out to be NP-complete, even for acyclic digraphs, as soon as the bound on the maximum out-degree of the support digraph is 2 or more. For non-acyclic digraphs, we show that even determining the existence of a flow $x$ of value $9$ where $D_x$ has maximum out-degree at most $2$ is NP-complete. In Section~\ref{sec:highconflow} we show that every 2-arc-strong $(s,t)$-flow network has a maximum flow whose support is also 2-arc-strong and show by a construction that this property does not generalize to higher arc-connectivities. The topic of Section~\ref{sec:fragile} is finding a maximum flow which is as secure as possible towards arc-deletions, that is, we want a maximum flow $D_x$ so that the value of a maximum flow in $D_x$ only drops by a small amount when we delete a few arcs.
In Section~\ref{sec:decompintopaths} we study how much flow we can send when we are only allowed to use $p$ paths for some integer $p$. Again we also consider the cases where these paths need to be arc-disjoint, respectively internally vertex-disjoint.  We show that in the case of acyclic networks, the arc-disjoint and  vertex-disjoint cases  can be solved in polynomial time by an XP-algorithm, parameterized by $p$, which is inspired by the solution in~\cite{fortuneTCS10} of the weak-$p$-linkage problem. We also show that, unless FPT=W[1]\footnote{For definitions of the problem classes FPT and W[1] we refer the reader to~\cite{cygan2015}.}, there can be no FPT-algorithm for this problem. In Section~\ref{sec:sepflow} we study the problem of finding a maximum flow which can be sent along paths so that no vertex is on more than $p$ paths. For every fixed $p\geq 1$, this turns out to be NP-hard already for acyclic digraphs and capacities 1 and 2.

\section{Preliminaries}\label{sec:prelim}

Generally notation follows~\cite{bang2009} so we only introduce a few things here for the convenience of the reader. The arc-set of a digraph $D$ is denoted by $A(D)$ and its vertex-set is denoted by $V(D)$. The {\bf maximum out-degree} and the {\bf maximum in-degree} of $D$, denoted respectively by $\Delta^+(D)$ and $\Delta^-(D)$, are defined respectively as $\max_{v\in V(D)} d^+(v)$ and $\max_{v\in V(D)} d^-(v)$. Given two vertices $s,t$ of a digraph $D$, we denote by $\lambda_{D}(s,t)$ the {\bf $(s,t)$-arc-connectivity} of $D$, that is the maximum number of pairwise arc-disjoint paths from $s$ to $t$ in $D$.

Let ${\cal N}= (D,s,t,c)$ be a flow  network and $x$ an $(s,t)$-flow
on $\cal N$. The {\bf value} of the flow $x$,
denoted by $|x|$, is the sum  of the $x$-values of the arcs going out from $s$, that is
$|x|= \sum_{u\in N^+(s)}x(us)$. The network $\cal N$ is {\bf acyclic} if $D$ is an
acyclic digraph.

Two very special flows are path-flows and cycle-flows.
A {\bf path-flow} (along $P$) is a flow $x$ such that for some integer $r$ we have $x(uv)=r$ for every arc $uv\in A(P)$ and $x(uv)=0$ if $uv\not\in A(P)$. A {\bf cycle-flow} (along $C$) is defined similarly with $P$ replaced by $C$ above.
The following very important property of flows is folklore. For a proof, see e.g.~\cite{bang2009}.

\begin{lemma}\label{lem:decompose-flow}
    Every flow $x$ in a network ${\cal N}=(D=(V,A),c)$ can be decomposed into (written as the arc-sum of) at most $|V|+|A|$ paths and cycle-flows such that at most $|A|$ of these are cycle-flows.
\end{lemma}

Let ${\cal N}= (D,s,t,c)$ be a flow network and $x$ an $(s,t)$-flow
on $\cal N$. The {\bf support digraph} of $x$, denoted by $D_x$, is
the digraph obtained from $D$ by removing all the arcs carrying no
flow, that is $A(D_x)=\{uv \mid uv \in A(D) \textrm{ and }
x(uv)>0\}$.

Lemma~\ref{lem:decompose-flow} immediately implies the following classical lemma which allows us to turn $D_x$ into an acyclic
digraph.

\begin{lemma}
\label{lem:acyclic-support}
Let $x$ be a flow on a network ${\cal N}=(D,s,t,c)$. There exists a
flow $x'$ on $\cal N$ with the same value as $x$ such that
$D_{x'}$ is acyclic. Moreover $D_{x'}$ is a subdigraph of $D_x$.
\end{lemma}

\begin{proof}
If $D_x$ is not acyclic, then let $C$ be a (directed) cycle of $D_x$. Notice that $C$ does not contain $s$, as we may assume that $s$ is a source in $D_x$. Let $u_0v_0$ be an arc of $C$ with minimum flow value along $C$.
And now, let $y$ be defined by $y(uv)=x(uv)$ if $uv$
is not an arc of $C$ and $y(uv)=x(uv)-x(u_0v_0)$ if $uv$ is an arc
of $C$. It is easy to check that $y$ is a flow on $\cal N$,
with the same value than $x$, as $C$ does not contain $s$, and that
$D_y$ is a subgraph of $D_x$ not containing $u_0v_0$. Repeating
this process of removing at least one arc from $D_x$ while it contains
a cycle stops with  a flow $x'$ with $D_{x'}$ being acyclic.
\end{proof}

The {\sc $p$-Linkage} problem is the following: Given a digraph $D$ and $2p$ distinct vertices $s_1,\ldots{},s_p, t_1,\ldots{},t_p$; decide if $D$ contains $p$ vertex-disjoint paths $P_1,\ldots{},P_p$ such that $P_i$ is an $(s_i,t_i)$-path for $i\in [p]$. The {\sc Weak-$p$-Linkage } problem is the same as above but where we only require that the paths are arc-disjoint.

\begin{theorem}[Fortune, Hopcroft and Wyllie~\cite{fortuneTCS10}]
    \label{thm:k_linkage}
    The {\sc $p$-Linkage} problem and the {\sc Weak-$p$-Linkage } problem are both NP-complete for every fixed $p\geq 2$. 
\end{theorem}

\begin{theorem}[Fortune, Hopcroft and Wyllie~\cite{fortuneTCS10}]
    \label{thm:k_linkage_acyclic}
    The {\sc $p$-Linkage} problem and the {\sc Weak-$p$-Linkage } problem are both solvable in polynomial time for every fixed $p\geq 1$ when the input is an acyclic digraph. If $p$ is part of the input, then both problems become NP-complete.
\end{theorem}

\begin{theorem}[Slivkins~\cite{slivkinsSJDM24}]
    \label{thm:k_linkage_W1_hard}
    The {\sc $p$-Linkage} problem and the {\sc Weak-$p$-Linkage } problem parameterised by $p$ are both W[1]-hard when the input is an acyclic digraph.
\end{theorem}

The {\sc $(3,B2)$-SAT} problem is the restriction of $3$-SAT in which every litteral appears exactly twice ({\it i.e.} every variable appears twice positively and twice negatively).

\begin{theorem}[{\cite[Theorem~1]{berman2004approximation}}]
    \label{thm:3_B2_SAT_NPc}
    {\sc $(3,B2)$-SAT} is NP-complete.
\end{theorem}

\section{Degree constrained flows}\label{sec:degreeconstrained}
\subsection{Bounded maximum out-degree}
In this section, we are concerned with the following problem.

\defproblem{\sc $(\Delta^+ \le k)$-Max-Flow}{A flow network ${\cal N}=(D,s,t,c)$.}{The maximum value of a flow $x$ on $\cal N$ such that $\Delta^+(D_x)\le k$.}

When $k=1$, the problem is known as the widest path problem or the maximum capacity path problem and is known to be polynomial time solvable by a modification of Dijkstra's shortest path algorithm, see e.g.~\cite[Excercise 4.37]{ahuja1993}. We show that the problem turns out to be NP-hard for every fixed $k\geq 2$ even on acyclic networks. Note that, in the case where all the capacity values are 1, the problem is solvable in polynomial time. Indeed, replace every vertex $x\neq t$ by two vertices $x^+$ and $x^-$, add an arc from $x^-$ to $x^+$ with capacity $k$ and the arcs from $N^-(x)$ to $x^-$ and the arcs from $x^+$ to $N^+(x)$ all with capacity 1. Then there is a one-to-one correspondence between the flows in the new network and the $(\Delta^+\le k)$-flows in the initial one.

\begin{theorem}\label{thm:Delta2acyclicflow}
    For every $k\geq 2$, {\sc $(\Delta^+ \le k)$-Max-Flow} is NP-hard even when restricted to acyclic networks.
\end{theorem}
\begin{proof}
We first prove the statement for $k=2$ and then explain how to modify the proof to prove it for all $k\geq 3$.
We will show how to reduce 3-SAT to the {\sc $(\Delta^+ \le 2)$-Max-Flow} problem on acyclic networks.
The first part of the reduction  we use was used in several papers, see e.g.~\cite{bangTCS526,bangTCS438} it is also similar to the reduction used in Section 4 of~\cite{evenSJC5}.

Let $W[p,q]$ be the digraph with vertices $\{u,v,y_1,y_2,\dots{}y_p,z_1,z_2,\ldots z_q\}$ and the arcs of the two $(u,v)$-paths $uy_1y_2\ldots{}y_pv$ and $uz_1z_2\ldots{}z_qv$. Let $H$ be the digraph on 4 vertices $a_1,a_2,a_3,y$ and arcs $a_1y, a_2y,a_3y$.
Let ${\cal F}$ be an instance of 3-SAT with variables $\varsymb_1,\varsymb_2,\ldots{},\varsymb_n$  and clauses $C_1,C_2,\ldots{},C_m$. By adding dummy clauses if necessary, we may assume that each variable $\varsymb_i$ occurs at least once both in the negated and in  non-negated forms in $\cal F$. The ordering of the clauses $C_1,C_2,\ldots{},C_m$ induces an ordering of  the occurrences of a variable $\varsymb_i$ and its negation $\bar{\varsymb}_i$ in these. With each variable $x=\varsymb_i$ we associate a copy of 
$W[p_i,q_i]$ where $\varsymb_i$ occurs $p_i$ times and $\bar{\varsymb}_i$ occurs $q_i$ times  in the clauses of $\cal F$. Identify end vertices of these digraphs by setting $v_i=u_{i+1}$ for $i=1,2,\ldots{},n-1$. 
Next, for each clause $C_i$ we take a copy $H_i$ of $H$ and identify the vertices $a_{i,1},a_{i,2},a_{i,3}$ of $H_i$ with vertices in the chain we build above as follows: assume $C_i$ contains variables $\varsymb_j,\varsymb_h,\varsymb_l$ (negated or not). If $\varsymb_j$ is not negated in $C_i$ and this is the $r$'th copy of $\varsymb_j$ (in the order of the clauses that use $\varsymb_j$), then we identify $a_{i,1}$ with $y_{j,r}$ and if $C_i$ contains $\bar{\varsymb}_j$ and this is the $r$'th occurrence of $\bar{\varsymb}_j$, then we identify $a_{i,1}$ with $z_{j,r}$. We make similar identifications for $a_{i,2},a_{i,3}$ with variables $\varsymb_h,\varsymb_\ell$. 

\begin{remark}
\label{rem:satisfiable}
It was shown in~\cite{bangTCS438} that ${\cal F}$ is satisfiable if and only if the digraph $\tilde{D}_{\cal F}$ constructed so far has a $(u_1,v_n)$-path which visits at least one vertex from each of the sets $\{a_{i,1},a_{i,2},a_{i,3}\}$, $i\in [m]$ (actually in~\cite{bangTCS438} the vertices $y_1,\ldots{},y_m$ were not included but they play no role in (this part of) the argument). 
\end{remark}

To finish the construction of $D_{\cal F}$ we add a new vertex $t$, the arc $v_nt$ and all the arcs $y_it$ for $i\in [m]$. Add a new vertex $s$ and the arc $su_1$. Finally we add the arcs $u_iv_i$ for $i\in [n]$.
This concludes the description of the digraph $D_{\cal F}$ with special vertices $s,t$. 

Now we construct a network ${\cal N}_{\cal F}=(D_{\cal F},s,t,c)$ by adding the following capacities to the arcs of $D_{\cal F}$:
\begin{itemize}
    \item the arc $su_1$ gets capacity $c(su_1)=3m+1$;
    \item the arc $v_nt$ gets capacity $c(v_nt)=2m+1$;
    \item the arcs of the form $u_iv_i$ get capacity $c(u_iv_i)=2m+1$;
    \item all arcs incident with $y_i$ get capacity 1 for $i\in [m]$;
    \item all remaining arcs get capacity $m$. 
\end{itemize}

\ifthenelse{\boolean{colouredfigures}}
{

\begin{figure}
    \begin{minipage}{\linewidth}
        \begin{center}	
          \begin{tikzpicture}[thick,scale=1, every node/.style={transform shape}]
            \tikzset{vertex/.style = {circle,fill=black,minimum size=4pt, inner sep=0pt}}
    	\tikzset{edge/.style = {->,> = latex'}}

            \node[vertex, label=left:$s$] (s) at (0.6,0) {};
            \node[vertex, label=above:$u_1$] (u1) at (2,0) {};
            \draw[edge,purple] (s) -- (u1) node [midway,yshift=0.5em] () {\scriptsize{$3m+1$}};
            
            \node[vertex, orange] (xT) at (3.5,1) {};
            \node[vertex,g-green] (xF1) at (3,-1) {};
            \node[vertex,g-blue] (xF2) at (4,-1) {};
            \node[vertex, label=above:$u_2$] (u2) at (5,0) {};
            
            \draw[edge,purple] (u1) -- (u2) node [midway,yshift=0.5em] () {\scriptsize{$2m+1$}};
            \draw[edge] (u1) -- (xT) node [midway,yshift=0.5em] () {\scriptsize{$m$}};
            \draw[edge] (u1) -- (xF1) node [midway,yshift=-0.5em,xshift=-0.15em] () {\scriptsize{$m$}};
            \draw[edge] (xF1) -- (xF2) node [midway,yshift=-0.5em] () {\scriptsize{$m$}};
            \draw[edge] (xF2) -- (u2) node [midway,yshift=-0.5em,xshift=0.15em] () {\scriptsize{$m$}};
            \draw[edge] (xT) -- (u2) node [midway,yshift=0.5em] () {\scriptsize{$m$}};

            \node[vertex,orange] (yT1) at (6,1) {};
            \node[vertex,g-green] (yT2) at (7,1) {};
            \node[vertex,g-blue] (yF) at (6.8,-1) {};
            \node[vertex, label=above:$u_3$] (u3) at (8,0) {};
            
            \draw[edge,purple] (u2) -- (u3) node [midway,yshift=0.5em] () {\scriptsize{$2m+1$}};
            \draw[edge] (u2) -- (yF) node [midway,yshift=-0.5em] () {\scriptsize{$m$}};
            \draw[edge] (u2) -- (yT1) node [midway,yshift=0.5em,xshift=-0.15em] () {\scriptsize{$m$}};
            \draw[edge] (yT1) -- (yT2) node [midway,yshift=0.5em] () {\scriptsize{$m$}};
            \draw[edge] (yT2) -- (u3) node [midway,yshift=0.5em,xshift=0.15em] () {\scriptsize{$m$}};
            \draw[edge] (yF) -- (u3) node [midway,yshift=-0.5em] () {\scriptsize{$m$}};

            \node[vertex,g-green] (zT) at (9.5,1) {};
            \node[vertex, orange] (zF1) at (9,-1) {};
            \node[vertex,g-blue] (zF2) at (10,-1) {};
            \node[vertex, label=above:$u_4$] (u4) at (11,0) {};
            
            \draw[edge,purple] (u3) -- (u4) node [midway,yshift=0.5em] () {\scriptsize{$2m+1$}};
            \draw[edge] (u3) -- (zT) node [midway,yshift=0.5em] () {\scriptsize{$m$}};
            \draw[edge] (u3) -- (zF1) node [midway,yshift=-0.5em,xshift=-0.15em] () {\scriptsize{$m$}};
            \draw[edge] (zF1) -- (zF2) node [midway,yshift=-0.5em] () {\scriptsize{$m$}};
            \draw[edge] (zF2) -- (u4) node [midway,yshift=-0.5em,xshift=0.15em] () {\scriptsize{$m$}};
            \draw[edge] (zT) -- (u4) node [midway,yshift=0.5em] () {\scriptsize{$m$}};

            \node[vertex, label=left:$y_1$, orange] (y1) at (3.2,-3) {};
            \node[vertex, label=left:$y_2$, g-green] (y2) at (5.8,-3) {};
            \node[vertex, label=left:$y_3$, g-blue] (y3) at (8.4,-3) {};
            
            \node[vertex, label=below:$t$] (t) at (5.8,-5.5) {};

            \draw[edge, orange] (xT) -- (y1);
            \draw[edge, orange] (yT1) -- (y1);
            \draw[edge, orange] (zF1) -- (y1);
            \draw[edge, orange] (y1) -- (t) node [midway,xshift=-0.6em] () {\scriptsize{$1$}};
            
            \draw[edge, g-green] (xF1) -- (y2);
            \draw[edge, g-green] (yT2) -- (y2);
            \draw[edge, g-green] (zT) -- (y2);
            \draw[edge, g-green] (y2) -- (t) node [midway,xshift=-0.6em] () {\scriptsize{$1$}};
            
            \draw[edge, g-blue] (xF2) -- (y3);
            \draw[edge, g-blue] (yF) -- (y3);
            \draw[edge, g-blue] (zF2) -- (y3);
            \draw[edge, g-blue] (y3) -- (t) node [midway,xshift=0.6em] () {\scriptsize{$1$}};
            
            \draw[edge, purple] (u4) to[in=0, out=-90] (t);
            \node[purple] (tcapacity) at (9,-5) {\scriptsize{$2m+1$}};
            
          \end{tikzpicture}
      \caption{The network $\mathcal{N}_\mathcal{F}$ when $\mathcal{F} = \textcolor{orange}{(x_1 \vee x_2 \vee \neg x_3)} \wedge \textcolor{g-green}{(\neg x_1 \vee x_2 \vee x_3)} \wedge \textcolor{g-blue}{(\neg x_1 \vee \neg x_2 \vee \neg x_3)}$.}
      \label{fig:D2_flow_acyclic}
    \end{center}    
  \end{minipage}
\end{figure}
}
{

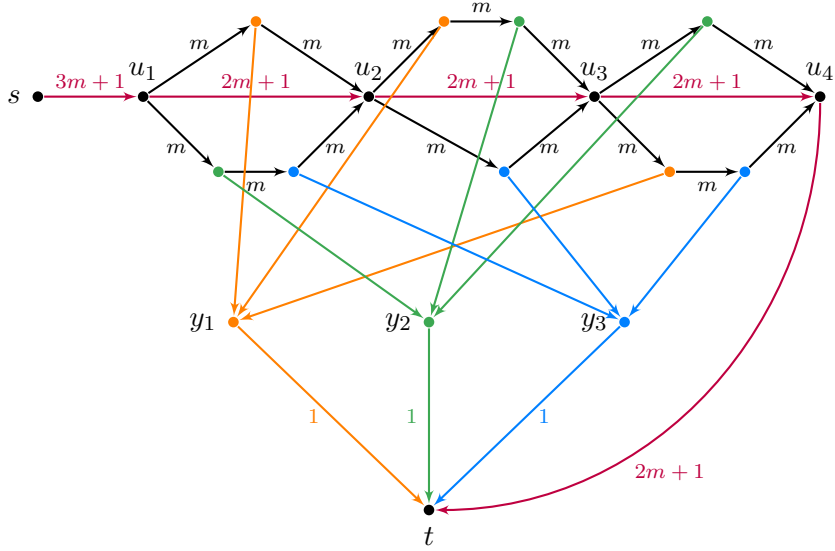
\begin{figure}
    \begin{minipage}{\linewidth}
        \begin{center}	
          \begin{tikzpicture}[thick,scale=1, every node/.style={transform shape}]
            \tikzset{vertex/.style = {circle,fill=black,minimum size=4pt, inner sep=0pt}}
    	\tikzset{edge/.style = {->,> = latex'}}

            \node[vertex, label=left:$s$] (s) at (0.6,0) {};
            \node[vertex, label=above:$u_1$] (u1) at (2,0) {};
            \draw[edge,dashed] (s) -- (u1) node [midway,yshift=0.5em] () {\scriptsize{$3m+1$}};
            
            \node[vertex] (xT) at (3.5,1) {};
            \node[vertex] (xF1) at (3,-1) {};
            \node[vertex] (xF2) at (4,-1) {};
            \node[vertex, label=above:$u_2$] (u2) at (5,0) {};
            
            \draw[edge,dashed] (u1) -- (u2) node [midway,yshift=0.5em] () {\scriptsize{$2m+1$}};
            \draw[edge] (u1) -- (xT) node [midway,yshift=0.5em] () {\scriptsize{$m$}};
            \draw[edge] (u1) -- (xF1) node [midway,yshift=-0.5em,xshift=-0.15em] () {\scriptsize{$m$}};
            \draw[edge] (xF1) -- (xF2) node [midway,yshift=-0.5em] () {\scriptsize{$m$}};
            \draw[edge] (xF2) -- (u2) node [midway,yshift=-0.5em,xshift=0.15em] () {\scriptsize{$m$}};
            \draw[edge] (xT) -- (u2) node [midway,yshift=0.5em] () {\scriptsize{$m$}};

            \node[vertex] (yT1) at (6,1) {};
            \node[vertex] (yT2) at (7,1) {};
            \node[vertex] (yF) at (6.8,-1) {};
            \node[vertex, label=above:$u_3$] (u3) at (8,0) {};
            
            \draw[edge,dashed] (u2) -- (u3) node [midway,yshift=0.5em] () {\scriptsize{$2m+1$}};
            \draw[edge] (u2) -- (yF) node [midway,yshift=-0.5em] () {\scriptsize{$m$}};
            \draw[edge] (u2) -- (yT1) node [midway,yshift=0.5em,xshift=-0.15em] () {\scriptsize{$m$}};
            \draw[edge] (yT1) -- (yT2) node [midway,yshift=0.5em] () {\scriptsize{$m$}};
            \draw[edge] (yT2) -- (u3) node [midway,yshift=0.5em,xshift=0.15em] () {\scriptsize{$m$}};
            \draw[edge] (yF) -- (u3) node [midway,yshift=-0.5em] () {\scriptsize{$m$}};

            \node[vertex] (zT) at (9.5,1) {};
            \node[vertex] (zF1) at (9,-1) {};
            \node[vertex] (zF2) at (10,-1) {};
            \node[vertex, label=above:$u_4$] (u4) at (11,0) {};
            
            \draw[edge,dashed] (u3) -- (u4) node [midway,yshift=0.5em] () {\scriptsize{$2m+1$}};
            \draw[edge] (u3) -- (zT) node [midway,yshift=0.5em] () {\scriptsize{$m$}};
            \draw[edge] (u3) -- (zF1) node [midway,yshift=-0.5em,xshift=-0.15em] () {\scriptsize{$m$}};
            \draw[edge] (zF1) -- (zF2) node [midway,yshift=-0.5em] () {\scriptsize{$m$}};
            \draw[edge] (zF2) -- (u4) node [midway,yshift=-0.5em,xshift=0.15em] () {\scriptsize{$m$}};
            \draw[edge] (zT) -- (u4) node [midway,yshift=0.5em] () {\scriptsize{$m$}};

            \node[vertex, label=left:$y_1$] (y1) at (3.2,-3) {};
            \node[vertex, label=left:$y_2$] (y2) at (5.8,-3) {};
            \node[vertex, label=left:$y_3$] (y3) at (8.4,-3) {};
            
            \node[vertex, label=below:$t$] (t) at (5.8,-5.5) {};

            \draw[edge, densely dotted] (xT) -- (y1);
            \draw[edge, densely dotted] (yT1) -- (y1);
            \draw[edge, densely dotted] (zF1) -- (y1);
            \draw[edge, densely dotted] (y1) -- (t) node [midway,xshift=-0.6em] () {\scriptsize{$1$}};
            
            \draw[edge, dash dot] (xF1) -- (y2);
            \draw[edge, dash dot] (yT2) -- (y2);
            \draw[edge, dash dot] (zT) -- (y2);
            \draw[edge, dash dot] (y2) -- (t) node [midway,xshift=-0.6em] () {\scriptsize{$1$}};
            
            \draw[edge, loosely dotted] (xF2) -- (y3);
            \draw[edge, loosely dotted] (yF) -- (y3);
            \draw[edge, loosely dotted] (zF2) -- (y3);
            \draw[edge, loosely dotted] (y3) -- (t) node [midway,xshift=0.6em] () {\scriptsize{$1$}};
            
            \draw[edge, dashed] (u4) to[in=0, out=-90] (t);
            \node[] (tcapacity) at (9,-5) {\scriptsize{$2m+1$}};
            
          \end{tikzpicture}
      \caption{The network $\mathcal{N}_\mathcal{F}$ when $\mathcal{F} = (x_1 \vee x_2 \vee \neg x_3) \wedge (\neg x_1 \vee x_2 \vee x_3) \wedge (\neg x_1 \vee \neg x_2 \vee \neg x_3)$.}
      \label{fig:D2_flow_acyclic}
    \end{center}    
  \end{minipage}
\end{figure}
}

Figure~\ref{fig:D2_flow_acyclic} illustrates the construction of $\mathcal{N}_{\mathcal{F}}$.
We claim that ${\cal N}_{\cal F}$ has an $(s,t)$-flow $x$ of value $3m+1$ such that $\Delta^+(D_x)\leq 2$ if and only if ${\cal F}$ is satisfiable. This follows from the remark above: suppose first that  ${\cal F}$ is satisfiable and let $\phi$ be a truth assignment satisfying ${\cal F}$. Then we can select for each clause $C_j$ the first literal $\ell_{j,r_j}$, $r_j\in [3]$ which is satisfied by $\phi$. By the remark, $D_{\cal F}$ has a $(u_1,v_n)$-path $P$ which contains all the vertices $a_{1,r_1},a_{2,r_2},\ldots{},a_{m,r_m}$. Now we obtain an $(s,t)$-flow $x$ of value $3m+1$ and maximum out-degree 2 in its support as follows:
\begin{itemize}
    \item send $2m+1$ units along the path $su_1u_2\ldots{}u_nt$
    \item use $P$ and the paths $a_{j,r_j}y_jt$ $j\in [m]$ to send one unit of flow from $s$ to each of the vertices $y_j$ and then to $t$ via the arc $y_jt$.
\end{itemize}

By construction $D_x$ has maximum out-degree 2.
To prove the other direction, suppose that $x$ is an $(s,t)$-flow of value $3m+1$ for which $D_x$ has maximum out-degree 2.
By the definition of capacities $x$ must send $2m+1$ units along the path $su_1u_2\ldots{}v_nt$. So as $\Delta^+(D_x)=2$ the remaining flow must be sent in a subdigraph $D'$ of $D_x$ so that each of $u_1,u_2,\ldots{},u_n$ have out-degree at most 1 in $D'$. It is easy to see that this implies the existence of a $(u_1,v_n)$-path in $\tilde{D}_{\cal F}$ which visits each of the sets $\{a_{i,1},a_{i,2},a_{i,3}\}$, $i\in [m]$ at least once. Hence, by the remark, ${\cal F}$ is satisfiable.

To modify the proof so that it works for any $k\geq 3$ we just need to replace the arcs of the form $u_iv_i$ with $k-1$ paths $u_ir_{i,1}v_i,u_ir_{i,2}v_i,\ldots{},u_ir_{i,k-1}v_i$ and let all arcs of these paths have capacity $2m+1$. Now it is easy to show that the new network has an $(s,t)$-flow $x$ of value $m+(k-1)(2m+1)$ whose support has maximum out-degree $k$ if and only if the 3-SAT formula ${\cal F}$ is satisfiable.
\end{proof}

\begin{problem}
Is there a way to approximate the {\sc $(\Delta^+ \le k)$-flow}
problem? In other words, if ${\cal N}$ has a flow $x^*$ satisfying $\Delta^+(D_{x^*}) \leq k$, is it possible to check in polynomial time that it has a flow $x$, $\Delta^+(D_{x}) \leq k$, such that $|x| \geq \rho(k) \cdot |x^*|$ for some function $\rho : \mathbb{N} \to ]0,1[$?
\end{problem}

If such a function $\rho : \mathbb{N} \to ]0,1[$ exists, the result of the next section implies that $\rho(2)\leq \frac{8}{9}$ (see Corollary~\ref{cor:inapx_dplus_2}). 
For a similar type of questions on degrees of spanning trees in graphs, see the paper~\cite{furerSODA3}.

\subsection{Finding a flow with value close to the maximum out-degree}
In this section, we take into account the value of the flow that we want to obtain
and, for $\ell \ge 0$, we define the next problem.

\defproblem{\sc $(\Delta^+ \le k)$-flow of value $k+\ell$}{A flow network ${\cal N}=(D,s,t,c)$.}{Does there exist a flow $x$ of value at least $k+\ell$ on $\cal N$ with $\Delta^+(D_x)\le k$?}

It is clear that {\sc $(\Delta^+ \le k)$-flow of value $k+\ell$} is polynomial for $\ell=0$, as it is equivalent as the existence of a flow of value $k$. First we will see that the problem remains solvable in polynomial time for $\ell=1$.

\begin{theorem}
\label{thm:dplus-plus-1}
The problem {\sc $(\Delta^+ \le k)$-flow of value $k+1$} is solvable in polynomial time.
\end{theorem}

\begin{proof}
    Let $\mathcal{N}=(D=(V,A),s,t,c)$ be an instance of {\sc $(\Delta^+ \le k)$-flow of value $k+1$}.
    We assume $2\leq k \leq n-1$ (where $n=|V(D)|$), the problem being clearly solvable in polynomial time when $k=1$ or $k\geq n$.
    We first check whether $\mathcal{N}$ admits an $(s,t)$-flow of value at least $k+1$. If this is not the case, $\mathcal{N}$ is clearly a negative instance. We can also assume that every vertex $v$ (and every arc $a$) of $D$ belongs to a path from $s$ to $t$, as otherwise no acyclic flow from $s$ to $t$ sends flow  through $v$ (or $a$) and we may delete it. A vertex $u$ of $D$ is an {\bf $(s,t)$-vertex separator} if there is no path from $s$ to $t$ in $D - u$. Assume first that $D$ does not contain any $(s,t)$-vertex separator. Then we transform  $\mathcal{N}$ as follows. Denote by $v_1,\dots ,v_p$ the out-neighbours of $s$ with $c(sv_i)\ge 2$. Then for $i=1,\dots ,p$ we build the network $\mathcal{N}_i$ by adding first a new source $s^\star$ to $D$ with an arc $s^{\star}s$ with capacity $k-1$ and an arc  $s^{\star}v_i$ with capacity $c(sv_i)$. We delete the arc $sv_i$, and then we do the usual vertex splitting operation (see e.g. \cite[Section 4.2.4]{bang2009}): every vertex $v\notin \{s^\star,s,t\}$ is replaced by two vertices $v^-,v^+$ and the arc $v^-v^+$ with capacity $k$, and every arc $uv \in A$ is replaced by the arc $u^+v^-$ and it preserves its capacity. We call the arcs of the kind  $v^-v^+$, the {\bf special arcs}  of the new network $\mathcal{N}_i$. We claim that $\mathcal{N}$ contains a flow $x$ with value $k+1$ and $\Delta^+(D_x)\le k$ if, and only if, there exists $i\in [p]$ such that $\mathcal{N}_i$ contains a flow of value $k+1$.
    
    Indeed, first assume that $\mathcal{N}_i$ contains a flow $x_i$ of value $k+1$ for some $i$. Then for each arc $uv$ of $D$ different from $sv_i$, we assign to it the flow $x_i(u^+v^-)$ (or $x_i(sv^-)$ if $u=s$ or $x_i(u^+t)$ if $v=t$). We finally put on $sv_i$ the flow $x_i(s^\star v_i)$. It is easy to check that we obtain a flow $x$ of value $k+1$ in $\mathcal{N}$. Let us see that $\Delta^+(D_x)\le k$. As every special arc of $D_i$ carries at most $k$ units of flow, it is clear that $d^+_{D_x}(v)\le k$ for every $v\in D- s$. Moreover, as $k+1$ units of flow leave $s^\star$ in $\mathcal{N}_i$ and $c(s^\star s)=k-1$, we have $x(xv_i)=x_i(s^\star v_i)\ge 2$. So, in $\cal N$, we have $x(sv_i)\ge 2$ and then $d^+_{D_x}(s)\le k$. Thus,  $x$ is a flow on 
    $\mathcal{N}$  with value $k+1$ and $\Delta^+(D_x)\le k$.
    
Conversely, assume that $\mathcal{N}$ contains a flow $x$ with value $k+1$ and $\Delta^+(D_x)\le k$. As $d^+_{D_x}(s)\le k$, there exists an out-neighbour $v_i$ of $s$ with $x(sv_i)\ge 2$. Then, we consider the network ${\cal N}_i$. Assume that it does not admit a flow of value $k+1$. 
By the Max Flow Min Cut Theorem (see e.g. \cite[Theorem 4.5.3]{bang2009}) this  means that  ${\cal N}_i$ has a cut $(X,\overline{X})$ with $s^\star \in X$,  $t\in \overline{X}$ and capacity at most $k$.  If some special arc $u^-u^+$ has $u^-\in X$ and $u^+\in \overline{X}$, then this is the only arc from $X$ to $\overline{X}$ (as it has capacity exactly $k$). 
 But this implies that $u$ is an $(s,t)$-vertex separator in $\cal N$, contradicting our assumption. So no special arc goes across the cut. Suppose first that $s\in \overline{X}$. Then we must have $v_i\in X$ as $c(s^\star v_i)\geq 2$ and $c(s^\star s)=k-1$.  But then $(X-s^\star{},\overline{X})$ is a $(v_i,t)$-cut of capacity at most 1 in $\cal N$, contradicting that at least two units of the flow $x$ go from $v_i$ to $t$ in $\cal N$. Thus we must have $s\in X$. Now define $X'$ to be the  subset of $V$ consisting  of $s$ and all the vertices $u\in V$ with $u^-,u^+\in X$. It is straightforward to see that $(X',\overline{X'})$ is an $(s,t)$-cut of $\cal N$ with the same capacity as  $(X,\overline{X})$, contradicting the existence of a flow with value $k+1$ in $\cal N$.
    
    There are at most $n$ networks ${\cal N}_i$, and performing a search for  a flow of value $k+1$ in ${\cal N}_i$ can be done in time $O(kn^2)=O(n^3)$ as we need at most $k+1$ augmenting paths. 
    So, in time $O(n^4)$ we can decide if there exists a $(\Delta^+ \le k)$-flow of value $k+1$ in a network with no $(s,t)$-vertex separator.

    \medskip

    Assume now that $\cal N$ contains at least one $(s,t)$-vertex separator, and consider any path $P$ from $s$ to $t$ in $D$. All the $(s,t)$-vertex separators of $D$ belong to $P$, and we can enumerate them $s_1,\dots{}, s_r$ according to the order in which they appear along $P$. We denote also $s$ by $s_0$ and $t$ by $s_{r+1}$. Finally, for $j=0,\dots ,r$ we denote by $V_j$ the set of vertices of $D$ lying on some path from $s_j$ to $s_{j+1}$ and by $D_j$ the digraph induced by $D$ on $V_j$.
    As every vertex and every arc of $D$ is on a path from $s$ to $t$, it is then not hard to see that the following holds. See Figure~\ref{fig:separator} for an illustration.
    
    \begin{itemize}
        \item $V=\bigcup_{1\le j\le r} V_j$  \item for $1\leq i<j\leq r+1$ the sets  $V_i$ and $V_j$ have a vertex in common only if $j=i+1$ and this vertex is $s_{i+1}$
        \item  for $1\leq i<j\leq r+1$ there is an arc from $V_i$ and $V_j$ only if $j=i+1$ and this arc is incident to $s_{i+1}$.
    \end{itemize}

    \begin{figure}[hbtp]
        \begin{center}	
              \begin{tikzpicture}[thick,scale=1, every node/.style={transform shape}]
                \tikzset{vertex/.style = {circle,fill=black,minimum size=4pt, inner sep=0pt}}
                \tikzset{littlevertex/.style = {circle,fill=black,minimum size=0pt, inner sep=0pt}}
    	    \tikzset{edge/.style = {->,> = latex'}}
                \tikzset{bigvertex/.style = {shape=circle,draw, minimum size=2em}}

                \node[vertex, label=left:$s$] (s) at (0,0){};
                \node[vertex, label=below:$s_1$] (s1) at (2.5,0) {};                
                \node[vertex, label=below:$s_2$] (s2) at (5,0) {};
                
                \node[vertex, label=below:$s_r$] (sr) at (10,0) {};
                
                \node[vertex, label=right:$t$] (t) at (12.5,0){};

                \draw[edge,dotted] (s) to[in=135, out=45] (s1){};
                \draw[edge,dotted] (s) to[in=-135, out=-45] (s1){};
                \node[text width=1cm] at (1.5,0) {$D_0$}; 
                
                \draw[edge,dotted] (s1) to[in=135, out=45] (s2){};
                \draw[edge,dotted] (s1) to[in=-135, out=-45] (s2){};
                \node[text width=1cm] at (4,0) {$D_1$}; 

                \draw[edge,dotted] (sr) to[in=135, out=45] (t){};
                \draw[edge,dotted] (sr) to[in=-135, out=-45] (t){};
                \node[text width=1cm] at (11.5,0) {$D_r$}; 
                \draw[dotted] (s2) -- (5.4,0.3);
                \draw[dotted] (s2) -- (5.4,-0.3);      
                \draw[edge,dotted] (9.6,0.3) -- (sr);
                \draw[edge,dotted] (9.6,-0.3) -- (sr);
                
              \end{tikzpicture}
          \caption{The structure of the network $\mathcal{N}$.}
          \label{fig:separator}
        \end{center}
    \end{figure}
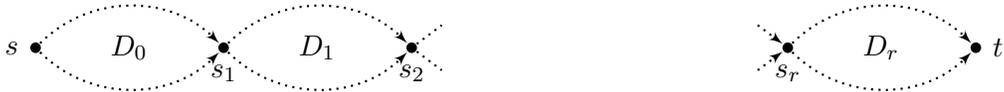
To conclude the proof, notice that, by construction, each $D_j$ has no $(s_j, s_{j+1})$-vertex separator. For $j=0,\dots ,r$, we denote by ${\cal N}_j$ the network $(D_j,s_j,s_{j+1},c_j)$  (where $c_j$ is $c$ restricted to  the arcs of $D_j$). It is clear that $\cal N$ contains a flow $x$ of  value $k + 1$ with $\Delta^+(D_x)\le k$ if, and only if, each ${\cal N}_j$ has such an $(s_j,s_{j+1})$-flow for every $j=0,\dots ,r$. So, by the previous case, denoting by $n_j$ the number of vertices of $D_j$, we can check in time $O(n_j^4)$ whether ${\cal N}_j$ contains such a flow for $j=0,\dots ,r$. Using the convexity of the function $z\rightarrow z^4$ and the fact that $\sum n_j\le 2n$, we conclude that it is possible to decide if $\cal N$ contains a flow $x$ with value $k + 1$ and $\Delta^+(D_x)\le k$ in time $O(n^4)$.
\end{proof}

We will now show that {\sc $(\Delta^+ \le k)$-flow of value $k+\ell$} is NP-complete for some fixed $k,\ell$, namely $k=2$ and $\ell =7$. Note that this is not a strengthening of Theorem~\ref{thm:Delta2acyclicflow} as we do not restrict to acyclic networks.

\begin{theorem}
\label{thm:npc_dplus}
The problem {\sc $(\Delta^+ \le 2)$-flow of value $9$} is NP-complete.
\end{theorem}

\begin{proof}
The following proof inspired by the proof of Theorem~9 in~\cite{Skut02}.The problem clearly being in NP, we prove the hardness by a reduction from {\sc $(3,B2)$-SAT}, which is NP-complete by Theorem~\ref{thm:3_B2_SAT_NPc}. Let $\mathcal{F} = (X,\mathcal{C})$ be an instance of {\sc $(3,B2)$-SAT}. We fix an arbitrary ordering $\varsymb_1,\dots, \varsymb_n$ on the variables $X$ and an ordering $C_1,\dots,C_m$ on the clauses $\mathcal{C}$.

We now describe the construction of a network $\mathcal{N}_{\mathcal{F}} = (D,s,t,c)$ such that $\mathcal{N}_{\mathcal{F}}$ admits a flow $x$ of value at least $9$ with $\Delta^+(D_x)\leq 2$ if and only if $\mathcal{F}$ is satisfiable.
Let $W$, our variable-gadget, be the digraph with vertex-set 
\[ V(W) = \{u,v\} \cup \{y_i,z_i \mid 1\leq i\leq 7\}\] 
and arc-set 
\[A(W) = \{uy_1, uz_1, uv, y_2y_4, z_2z_4,y_5y_7,z_5z_7,y_7v,z_7v\} \cup \{y_iy_{i+1},z_iz_{i+1} \mid 1\leq i \leq 6\},\]
see Figure~\ref{fig:vertex_gadget_npc_dplus} for an illustration and for the capacity $c(a)$ of every arc $a\in A(W)$. 

\begin{figure}[hbtp]
        \begin{center}	
              \begin{tikzpicture}[thick,scale=1, every node/.style={transform shape}]
                \tikzset{vertex/.style = {circle,fill=black,minimum size=4pt, inner sep=0pt}}
                \tikzset{littlevertex/.style = {circle,fill=black,minimum size=0pt, inner sep=0pt}}
    	    \tikzset{edge/.style = {->,> = latex'}}
         
                \node[vertex, label=left:$u$] (u) at (0,0){};
                \node[vertex, label=right:$v$] (v) at (9,0){};
                \foreach \i in {1,2,4,5,7}{
                    \node[vertex, label=above:$y_{\i}$] (y\i) at (\i+0.5, 1.7) {};
                    \node[vertex, label=below:$z_{\i}$] (z\i) at (\i+0.5, -1.7) {};
                }
                \foreach \i in {3,6}{
                    \node[vertex, label=above:$y_{\i}$] (y\i) at (\i+0.5, 2.4) {};
                    \node[vertex, label=below:$z_{\i}$] (z\i) at (\i+0.5, -2.4) {};
                }
                \foreach \i in {1,4}{
                    \pgfmathtruncatemacro{\j}{\i+1}
                    \draw[edge] (y\i) -- (y\j) node [midway,yshift=-0.6em] () {\scriptsize{$4$}};
                    \draw[edge] (z\i) -- (z\j) node [midway,yshift=0.6em] () {\scriptsize{$4$}};
                }
                \foreach \i in {2,3,5,6}{
                    \pgfmathtruncatemacro{\j}{\i+1}
                    \draw[edge] (y\i) -- (y\j) node [midway,yshift=0.5em] () {\scriptsize{$2$}};
                    \draw[edge] (z\i) -- (z\j) node [midway,yshift=-0.5em] () {\scriptsize{$2$}};
                }
                \foreach \i in {2,5}{
                    \pgfmathtruncatemacro{\j}{\i+2}
                    \draw[edge] (y\i) -- (y\j) node [midway,yshift=-0.6em] () {\scriptsize{$2$}};
                    \draw[edge] (z\i) -- (z\j) node [midway,yshift=0.6em] () {\scriptsize{$2$}};
                }
                \draw[edge] (u) -- (v) node [midway,yshift=0.6em] () {\scriptsize{$4$}};
                \draw[edge] (u) -- (y1) node [midway, xshift=-0.4em, yshift=0.4em] () {\scriptsize{$4$}};
                \draw[edge] (u) -- (z1) node [midway, xshift=-0.4em, yshift=-0.4em] () {\scriptsize{$4$}};
                \draw[edge] (y7) -- (v) node [midway, xshift=0.4em, yshift=0.4em] () {\scriptsize{$4$}};
                \draw[edge] (z7) -- (v) node [midway, xshift=0.4em, yshift=-0.4em] () {\scriptsize{$4$}};
                
              \end{tikzpicture}
            \caption{The variable-gadget $W$.}
          \label{fig:vertex_gadget_npc_dplus}
        \end{center}
    \end{figure}
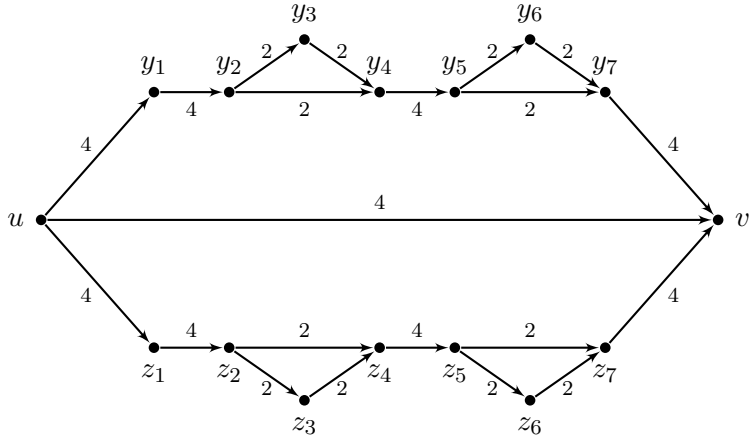

For every variable $\varsymb_i \in X$ we add a distinct copy $W^i$ of $W$ (preserving the capacities of $W$). For every vertex $w\in W$, we let $w^i$ designate the copy of $w$ in $W^i$. For every $i\in [n-1]$, we add the arc $v^iu^{i+1}$ with capacity $8$.

For every clause $C_j \in \mathcal{C}$, we add two vertices $q_j$ and $r_j$, and for every $j\in \{1,\dots,m-1\}$ we add the arc $r_jq_{j+1}$ with capacity $1$. For every variable $\varsymb_h$ appearing positively in $C_i, C_j$, $i<j$, and negatively in $C_k, C_\ell$, $k< \ell$, we add the following set of arcs with capacity $1$:
\[ \big\{q_iy_4^h, y_5^hr_i, q_jy_1^h, y_2^hr_j, q_kz_4^h, z_5^hr_k, q_\ell z_1^h, z_2^hr_\ell \big\}  \]

We finally add the source $s$, the sink $t$, the arcs $su^1,v^nt$ with capacity $8$, and the arcs $sq_1,r_mt$ with capacity $1$. We let $\mathcal{N}_{\mathcal{F}}=(D,s,t,c)$ be the obtained network, see Figure~\ref{fig:reduction_npc_dplus} for an illustration.

\ifthenelse{\boolean{colouredfigures}}
    {
\begin{figure}[hbtp]
        \begin{center}	
              \begin{tikzpicture}[scale=1, every node/.style={transform shape}]
                \tikzset{vertex/.style = {circle,fill=black,minimum size=4pt, inner sep=0pt}}
                \tikzset{littlevertex/.style = {circle,fill=black,minimum size=0pt, inner sep=0pt}}
    	    \tikzset{edge/.style = {->,> = latex'}}
         
                \node[vertex, label=left:$s$] (s) at (0,-1){};
                \node[vertex, label=right:$t$] (t) at (12.6,-1){};
                
                \node[vertex, label=above:$u^1$] (u1) at (1,1){};
                \node[vertex, label=above:$v^1$] (v1) at (1.7,1){};
                
                \draw (u1) to[out=60, in=120] (v1) {};
                \draw (u1) to (v1) {};
                \draw (u1) to[out=-60, in=-120] (v1) {};
                \node[vertex, label=above:$u^h$] (uh) at (3.8,1){};
                
                \draw[edge] (v1) -- (2.4,1) node [midway, yshift=-0.6em] () {\scriptsize 8}; 
                \draw[edge] (s) -- (u1) node [midway, xshift=-0.4em, yshift=0.4em] () {\scriptsize 8};
                \draw[dotted] (2.4,1) -- (3.1,1);
                \draw[edge] (3.1,1) -- (uh) node [midway, yshift=-0.6em] () {\scriptsize 8}; 
                
                \node[vertex,purple] (y1) at (4.5,1.7) {};
                \node[vertex,purple] (y2) at (5.4,1.7) {};
                \node[vertex] (y3) at (5.85,2) {};
                \node[vertex,orange] (y4) at (6.3,1.7) {};
                \node[vertex,orange] (y5) at (7.2,1.7) {};
                \node[vertex] (y6) at (7.65,2) {};
                \node[vertex] (y7) at (8.1,1.7) {};
                \node[vertex,g-green] (z1) at (4.5,0.3) {};
                \node[vertex,g-green] (z2) at (5.4,0.3) {};
                \node[vertex] (z3) at (5.85,0) {};
                \node[vertex, g-blue] (z4) at (6.3,0.3) {};
                \node[vertex, g-blue] (z5) at (7.2,0.3) {};
                \node[vertex] (z6) at (7.65,0) {};
                \node[vertex] (z7) at (8.1,0.3) {};
                \foreach \i in {1,2,3,4,5,6}{
                    \pgfmathtruncatemacro{\j}{\i+1}
                    \draw[edge] (y\i) to (y\j);
                    \draw[edge] (z\i) to (z\j);
                }
                \foreach \i in {2,5}{
                    \pgfmathtruncatemacro{\j}{\i+2}
                    \draw[edge] (y\i) to (y\j);
                    \draw[edge] (z\i) to (z\j);
                }
                \node[vertex, label=above:$v^h$] (vh) at (8.8,1){};
                \draw[edge] (uh) to (y1);
                \draw[edge] (uh) to (z1);
                \draw[edge] (y7) to (vh);
                \draw[edge] (z7) to (vh);
                
                \node[vertex, label=above:$u^n$] (un) at (10.9,1){};
                \node[vertex, label=above:$v^n$] (vn) at (11.6,1){};
                
                \draw (un) to[out=60, in=120] (vn) {};
                \draw (un) to (vn) {};
                \draw (un) to[out=-60, in=-120] (vn) {};
                
                \draw[edge] (10.2,1) -- (un)  node [midway, yshift=-0.6em] () {\scriptsize 8}; 
                \draw[edge] (vn) -- (t) node [midway, xshift=0.4em, yshift=0.4em] () {\scriptsize 8};
                \draw[dotted] (10.2,1) -- (9.5,1);
                \draw[edge] (vh) -- (9.5,1)node [midway, yshift=-0.6em] () {\scriptsize 8}; 

                \begin{scope}[xshift=1cm, yshift=-3cm]
                    \node[vertex, label=below:$q_1$] (q1) at (0,0){};
                    \draw[dashed, edge] (q1) to (72:0.8);
                    \draw[dashed, edge] (q1) to (90:0.8);
                    \draw[dashed, edge] (q1) to (54:0.8);
                \end{scope}
                \begin{scope}[xshift=11.6cm, yshift=-3cm]
                \node[vertex, label=below:$r_m$] (rm) at (0,0){};
                    \draw[dashed, edge] (126:0.8) to (rm);
                    \draw[dashed, edge] (90:0.8) to (rm);
                    \draw[dashed, edge] (108:0.8) to (rm);
                \end{scope}
                \draw[edge] (s) -- (q1) node [midway, xshift=-0.4em, yshift=-0.4em] () {\scriptsize 1};

                \begin{scope}[xshift=-1.7cm]
                \draw[dotted] (3.2,-3) -- (3.6,-3);
                
                \begin{scope}[xshift=4.3cm, yshift=-3cm]
                    \node[vertex, g-blue, label=below:$q_k$] (qk) at (0,0){};
                    \draw[dashed, edge] (qk) to (108:0.8);
                    \draw[dashed, edge] (qk) to (90:0.8);
                    \draw[g-blue, edge] (qk) -- (z4){};
                \end{scope}
                \begin{scope}[xshift=4.8cm, yshift=-3cm]
                    \node[vertex, g-blue, label=below:$r_k$] (rk) at (0,0){};
                    \draw[dashed, edge] (72:0.8) to (rk);
                    \draw[dashed, edge] (90:0.8) to (rk);
                    \draw[g-blue, edge] (z5) -- (rk){};
                \end{scope}
                \draw[edge] (3.6,-3) -- (qk) node [midway, yshift=0.6em] () {\scriptsize 1};
                \draw[dotted] (3.2,-3) -- (3.6,-3);
                \draw[edge] (rk) -- (5.5,-3) node [midway, yshift=0.6em] () {\scriptsize 1};
                \draw[dotted] (5.5,-3) -- (5.9,-3);

                \begin{scope}[xshift=6.6cm, yshift=-3cm]
                    \node[vertex, g-green, label=below:$q_\ell$] (ql) at (0,0){};
                    \draw[dashed, edge] (ql) to (108:0.8);
                    \draw[dashed, edge] (ql) to (90:0.8);
                    \draw[g-green, edge] (ql) -- (z1){};
                \end{scope}
                \begin{scope}[xshift=7.1cm, yshift=-3cm]
                    \node[vertex, g-green, label=below:$r_\ell$] (rl) at (0,0){};
                    \draw[dashed, edge] (72:0.8) to (rl);
                    \draw[dashed, edge] (108:0.8) to (rl);
                    \draw[g-green, edge] (z2) -- (rl){};
                \end{scope}
                \draw[edge] (5.9,-3) -- (ql) node [midway, yshift=0.6em] () {\scriptsize 1};
                \draw[edge] (rl) -- (7.8,-3) node [midway, yshift=0.6em] () {\scriptsize 1};
                \draw[dotted] (7.8,-3) -- (8.2,-3);

                \begin{scope}[xshift=8.9cm, yshift=-3cm]
                    \node[vertex, orange, label=below:$q_i$] (qi) at (0,0){};
                    \draw[dashed, edge] (qi) to (108:0.8);
                    \draw[dashed, edge] (qi) to (90:0.8);
                    \draw[orange, edge] (qi) -- (y4){};
                \end{scope}
                \begin{scope}[xshift=9.4cm, yshift=-3cm]
                    \node[vertex, orange, label=below:$r_i$] (ri) at (0,0){};
                    \draw[dashed, edge] (72:0.8) to (ri);
                    \draw[dashed, edge] (90:0.8) to (ri);
                    \draw[orange, edge] (y5) -- (ri){};
                \end{scope}
                \draw[edge] (8.2,-3) -- (qi) node [midway, yshift=0.6em] () {\scriptsize 1};
                \draw[edge] (ri) -- (10.1,-3) node [midway, yshift=0.6em] () {\scriptsize 1};
                \draw[dotted] (10.1,-3) -- (10.5,-3);
                
                \begin{scope}[xshift=11.2cm, yshift=-3cm]
                    \node[vertex, purple, label=below:$q_j$] (qj) at (0,0){};
                    \draw[dashed, edge] (qj) to (108:0.8);
                    \draw[dashed, edge] (qj) to (90:0.8);
                    \draw[purple, edge] (qj) -- (y1){};
                \end{scope}
                \begin{scope}[xshift=11.7cm, yshift=-3cm]
                    \node[vertex, purple, label=below:$r_j$] (rj) at (0,0){};
                    \draw[dashed, edge] (72:0.8) to (rj);
                    \draw[dashed, edge] (90:0.8) to (rj);
                    \draw[purple, edge] (y2) -- (rj){};
                \end{scope}
                \draw[edge] (10.5,-3) -- (qj) node [midway, yshift=0.6em] () {\scriptsize 1};
                \draw[edge] (rj) -- (12.4,-3) node [midway, yshift=0.6em] () {\scriptsize 1};
                \draw[dotted] (12.4,-3) -- (12.8,-3);
                \draw[edge] (rm) -- (t) node [midway, xshift=0.4em, yshift=-0.4em] () {\scriptsize 1};
                \end{scope}
                \draw[edge] (uh) to (vh);
              \end{tikzpicture}
            \caption{An illustration of the network $\mathcal{N}_{\mathcal{F}}$. Three undirected edges between the same pair of vertices represent a copy of $W$. Arcs in a copy of $W$ have the same capacities as in Figure~\ref{fig:vertex_gadget_npc_dplus}. Arcs incident to vertices in $\{q_i,r_i \mid 1 \leq i \leq m\}$ have capacity $1$.}
          \label{fig:reduction_npc_dplus}
        \end{center}
    \end{figure}
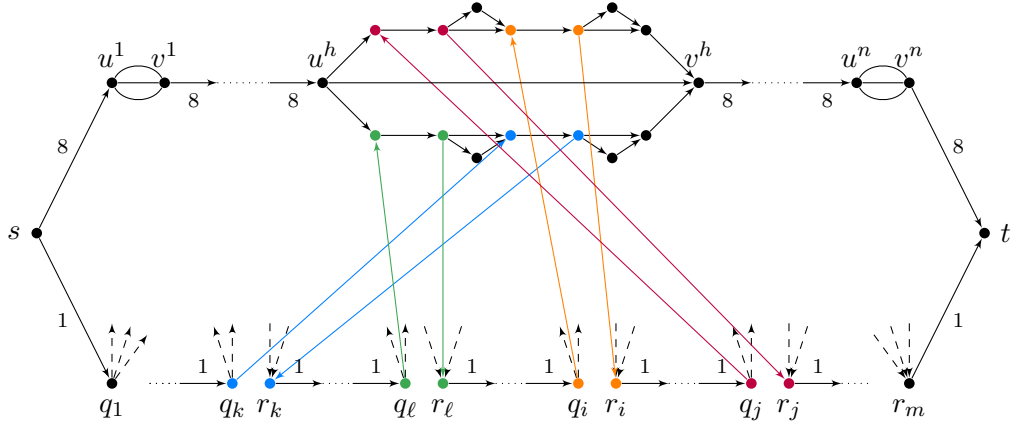
    }
    {
\begin{figure}[hbtp]
        {\scriptsize
        \begin{center}	
              \begin{tikzpicture}[scale=1, every node/.style={transform shape}]
                \tikzset{vertex/.style = {circle,fill=black,minimum size=4pt, inner sep=0pt}}
                \tikzset{littlevertex/.style = {circle,fill=black,minimum size=0pt, inner sep=0pt}}
    	    \tikzset{edge/.style = {->,> = latex'}}
         
                \node[vertex, label=left:$s$] (s) at (0,-1){};
                \node[vertex, label=right:$t$] (t) at (12.6,-1){};
                
                \node[vertex, label=above:$u^1$] (u1) at (1,1){};
                \node[vertex, label=above:$v^1$] (v1) at (1.7,1){};
                
                \draw (u1) to[out=60, in=120] (v1) {};
                \draw (u1) to (v1) {};
                \draw (u1) to[out=-60, in=-120] (v1) {};
                \node[vertex, label=above:$u^h$] (uh) at (3.8,1){};
                
                \draw[edge] (v1) -- (2.4,1) node [midway, yshift=-0.6em] () {\scriptsize 8}; 
                \draw[edge] (s) -- (u1) node [midway, xshift=-0.4em, yshift=0.4em] () {\scriptsize 8};
                \draw[dotted] (2.4,1) -- (3.1,1);
                \draw[edge] (3.1,1) -- (uh) node [midway, yshift=-0.6em] () {\scriptsize 8}; 
                
                \node[vertex] (y1) at (4.5,1.7) {};
                \node[vertex] (y2) at (5.4,1.7) {};
                \node[vertex] (y3) at (5.85,2) {};
                \node[vertex] (y4) at (6.3,1.7) {};
                \node[vertex] (y5) at (7.2,1.7) {};
                \node[vertex] (y6) at (7.65,2) {};
                \node[vertex] (y7) at (8.1,1.7) {};
                \node[vertex] (z1) at (4.5,0.3) {};
                \node[vertex] (z2) at (5.4,0.3) {};
                \node[vertex] (z3) at (5.85,0) {};
                \node[vertex] (z4) at (6.3,0.3) {};
                \node[vertex] (z5) at (7.2,0.3) {};
                \node[vertex] (z6) at (7.65,0) {};
                \node[vertex] (z7) at (8.1,0.3) {};
                \foreach \i in {1,2,3,4,5,6}{
                    \pgfmathtruncatemacro{\j}{\i+1}
                    \draw[edge] (y\i) to (y\j);
                    \draw[edge] (z\i) to (z\j);
                }
                \foreach \i in {2,5}{
                    \pgfmathtruncatemacro{\j}{\i+2}
                    \draw[edge] (y\i) to (y\j);
                    \draw[edge] (z\i) to (z\j);
                }
                \node[vertex, label=above:$v^h$] (vh) at (8.8,1){};
                \draw[edge] (uh) to (y1);
                \draw[edge] (uh) to (z1);
                \draw[edge] (y7) to (vh);
                \draw[edge] (z7) to (vh);
                
                \node[vertex, label=above:$u^n$] (un) at (10.9,1){};
                \node[vertex, label=above:$v^n$] (vn) at (11.6,1){};
                
                \draw (un) to[out=60, in=120] (vn) {};
                \draw (un) to (vn) {};
                \draw (un) to[out=-60, in=-120] (vn) {};
                
                \draw[edge] (10.2,1) -- (un)  node [midway, yshift=-0.6em] () {\scriptsize 8}; 
                \draw[edge] (vn) -- (t) node [midway, xshift=0.4em, yshift=0.4em] () {\scriptsize 8};
                \draw[dotted] (10.2,1) -- (9.5,1);
                \draw[edge] (vh) -- (9.5,1)node [midway, yshift=-0.6em] () {\scriptsize 8}; 

                \begin{scope}[xshift=1cm, yshift=-3cm]
                    \node[vertex, label=below:$q_1$] (q1) at (0,0){};
                    \draw[dashed, edge] (q1) to (72:0.8);
                    \draw[dashed, edge] (q1) to (90:0.8);
                    \draw[dashed, edge] (q1) to (54:0.8);
                \end{scope}
                \begin{scope}[xshift=11.6cm, yshift=-3cm]
                \node[vertex, label=below:$r_m$] (rm) at (0,0){};
                    \draw[dashed, edge] (126:0.8) to (rm);
                    \draw[dashed, edge] (90:0.8) to (rm);
                    \draw[dashed, edge] (108:0.8) to (rm);
                \end{scope}
                \draw[edge] (s) -- (q1) node [midway, xshift=-0.4em, yshift=-0.4em] () {\scriptsize 1};

                \begin{scope}[xshift=-1.7cm]
                \draw[dotted] (3.2,-3) -- (3.6,-3);
                
                \begin{scope}[xshift=4.3cm, yshift=-3cm]
                    \node[vertex, label=below:$q_k$] (qk) at (0,0){};
                    \draw[dashed, edge] (qk) to (108:0.8);
                    \draw[dashed, edge] (qk) to (90:0.8);
                    \draw[edge] (qk) -- (z4){};
                \end{scope}
                \begin{scope}[xshift=4.8cm, yshift=-3cm]
                    \node[vertex, label=below:$r_k$] (rk) at (0,0){};
                    \draw[dashed, edge] (72:0.8) to (rk);
                    \draw[dashed, edge] (90:0.8) to (rk);
                    \draw[edge] (z5) -- (rk){};
                \end{scope}
                \draw[edge] (3.6,-3) -- (qk) node [midway, yshift=0.6em] () {\scriptsize 1};
                \draw[dotted] (3.2,-3) -- (3.6,-3);
                \draw[edge] (rk) -- (5.5,-3) node [midway, yshift=0.6em] () {\scriptsize 1};
                \draw[dotted] (5.5,-3) -- (5.9,-3);

                \begin{scope}[xshift=6.6cm, yshift=-3cm]
                    \node[vertex, label=below:$q_\ell$] (ql) at (0,0){};
                    \draw[dashed, edge] (ql) to (108:0.8);
                    \draw[dashed, edge] (ql) to (90:0.8);
                    \draw[edge] (ql) -- (z1){};
                \end{scope}
                \begin{scope}[xshift=7.1cm, yshift=-3cm]
                    \node[vertex, label=below:$r_\ell$] (rl) at (0,0){};
                    \draw[dashed, edge] (72:0.8) to (rl);
                    \draw[dashed, edge] (108:0.8) to (rl);
                    \draw[edge] (z2) -- (rl){};
                \end{scope}
                \draw[edge] (5.9,-3) -- (ql) node [midway, yshift=0.6em] () {\scriptsize 1};
                \draw[edge] (rl) -- (7.8,-3) node [midway, yshift=0.6em] () {\scriptsize 1};
                \draw[dotted] (7.8,-3) -- (8.2,-3);

                \begin{scope}[xshift=8.9cm, yshift=-3cm]
                    \node[vertex, label=below:$q_i$] (qi) at (0,0){};
                    \draw[dashed, edge] (qi) to (108:0.8);
                    \draw[dashed, edge] (qi) to (90:0.8);
                    \draw[edge] (qi) -- (y4){};
                \end{scope}
                \begin{scope}[xshift=9.4cm, yshift=-3cm]
                    \node[vertex, label=below:$r_i$] (ri) at (0,0){};
                    \draw[dashed, edge] (72:0.8) to (ri);
                    \draw[dashed, edge] (90:0.8) to (ri);
                    \draw[edge] (y5) -- (ri){};
                \end{scope}
                \draw[edge] (8.2,-3) -- (qi) node [midway, yshift=0.6em] () {\scriptsize 1};
                \draw[edge] (ri) -- (10.1,-3) node [midway, yshift=0.6em] () {\scriptsize 1};
                \draw[dotted] (10.1,-3) -- (10.5,-3);
                
                \begin{scope}[xshift=11.2cm, yshift=-3cm]
                    \node[vertex, label=below:$q_j$] (qj) at (0,0){};
                    \draw[dashed, edge] (qj) to (108:0.8);
                    \draw[dashed, edge] (qj) to (90:0.8);
                    \draw[edge] (qj) -- (y1){};
                \end{scope}
                \begin{scope}[xshift=11.7cm, yshift=-3cm]
                    \node[vertex, label=below:$r_j$] (rj) at (0,0){};
                    \draw[dashed, edge] (72:0.8) to (rj);
                    \draw[dashed, edge] (90:0.8) to (rj);
                    \draw[edge] (y2) -- (rj){};
                \end{scope}
                \draw[edge] (10.5,-3) -- (qj) node [midway, yshift=0.6em] () {\scriptsize 1};
                \draw[edge] (rj) -- (12.4,-3) node [midway, yshift=0.6em] () {\scriptsize 1};
                \draw[dotted] (12.4,-3) -- (12.8,-3);
                \draw[edge] (rm) -- (t) node [midway, xshift=0.4em, yshift=-0.4em] () {\scriptsize 1};
                \end{scope}
                \draw[edge] (uh) to (vh);
              \end{tikzpicture}
            \caption{An illustration of the network $\mathcal{N}_{\mathcal{F}}$. Three undirected edges between the same pair of vertices represent a copy of $W$. Arcs in a copy of $W$ have the same capacities as in Figure~\ref{fig:vertex_gadget_npc_dplus}. Arcs incident to vertices in $\{q_i,r_i \mid 1 \leq i \leq m\}$ have capacity $1$.}
          \label{fig:reduction_npc_dplus}
        \end{center}
        }
    \end{figure}
    }

    We now prove the equivalence between the instances $\mathcal{F}$ and $\mathcal{N}_{\mathcal{F}}$. Assume first that $\mathcal{F}$ is satisfied by a truth assignment $\phi$, we will show that $\mathcal{N}_{\mathcal{F}}$ admits a flow $x$ of value $9$ such that $\Delta^+(D_x) \leq 2$. For every $j\in [m]$, we fix an arbitrary variable $\varsymb_{\iota(j)}$ such that $C_j$ is satisfied by $\varsymb_{\iota(j)}$ or its negation in $\phi$. If $\varsymb_{\iota(j)}$ satisfies $C_j$, we let $w_j$ be the out-neighbour of $q_j$ in $\{y_1^{\iota(j)},y_4^{\iota(j)}\}$. Else if $\bar{\varsymb}_{\iota(j)}$ satisfies $C_j$, we let $w_j$ be the out-neighbour of $q_j$ in $\{z_1^{\iota(j)},z_4^{\iota(j)}\}$. In both cases, we let $w_j'$ be the unique out-neighbour of $w_j$ (observe that $w_j'$ is an in-neighbour of $r_j$).
    We define  $x$ as follows:
    \begin{itemize}
        \item $x(su^1) = x(v^nt) = 8$;
        \item $x(sq_1) = x(r_mt) = 1$;
        \item $\forall i\in [n]$, $x(u^iv^{i}) = 4$;
        \item $\forall i\in [n-1]$, $x(v^iu^{i+1}) = 8$;
        \item $\forall j\in [m-1]$, $x(r^jq^{j+1}) = 1$;
        \item $\forall i\in [n]$ such that $\phi(\varsymb_i)$ is true, $x(u^iz_1^i) = x(z_1^iz_2^i) = x(z_4^iz_5^i)= x(z_7^iv^i) = 4$ and $x(z_2^iz_3^i) = x(z_3^iz_4^i) = x(z_2^iz_4^i) = x(z_5^iz_6^i)= x(z_6^iz_7^i)= x(z_5^iz_7^i) = 2$;
        \item $\forall i\in [n]$ such that $\phi(\varsymb_i)$ is false, $x(u^iy_1^i) = x(y_1^iy_2^i) = x(y_4^iy_5^i)= x(y_7^iv^i) = 4$ and $x(y_2^iy_3^i)= x(y_3^iy_4^i) = x(y_2^iy_4^i) = x(y_5^iy_6^i)= x(y_6^iy_7^i)= x(y_5^iy_7^i) = 2$; and
        \item $\forall j\in [m]$, $x(q_jw_j) = x(w_jw_j') = x(w_j'r_j) = 1$. 
    \end{itemize}

    By definition, $x$ is a flow of value $9$. We now prove that $\Delta^+(D_x) \leq 2$. By construction, the only vertices that may have out-degree more than $2$ in $D_x$ are vertices in $\{y_2^i,y_5^i,z_2^i,z_5^i\mid i\in [n]\}$ identified as $w_j'$ for some $j\in [m]$. Assume that, for some indices $i,j$, $w_j'$ belongs to $\{y_2^i,y_5^i,z_2^i,z_5^i\}$. By choice of $w_j$ and $w_j'$, either $C_j$ is satisfied by $\varsymb_i$ and $w_j'\in \{z_2^i,z_5^i\}$ or $C_j$ is satisfied by $\bar{\varsymb}_i$ and $w_j'\in \{y_2^i,y_5^i\}$. In both cases, by definition of $x$, we conclude that $w_j'$ has out-degree $1$ in $D_x$. Hence $\Delta^+(D_x)\leq 2$ as desired.

    \medskip

    Conversely, assume now that $\mathcal{N}_{\mathcal{F}}$ admits a flow $y$ of value $9$ such that $\Delta^+(D_y) \leq 2$. Among all such flows $y$, we choose $x$ for which $|A(D_x)|$ is minimum.
    The two following claims show that $x$ satisfies some structural properties.
    \begin{claim}
        \label{claim:structure_x_9_side}
        For every $i\in [n]$, $x(u^iv^i) = 4$ and $x(v^{i-1}u^i) =8$, where $v^0$ is identified as $s$. Moreover, $x(u^iy_1^i) =  x(y_7^iv^i)$ and $x(u^iz_1^i) = x(z_7^iv^i)$.
    \end{claim}
    \begin{subproof}
        We proceed by induction on $i\in [n]$. Since $|x| = 9$, we necessarily have $x(su^1) = 8$.
        
        Assume now that, for some $i\in [n]$, $x(v^{i-1}u^i) =8$. Since $\Delta^+(D_x) \leq 2$, and because $8$ units of flow enter $u^i$, exactly two arcs of $\{u^iv^i,u^iy_1^i,u^iz_1^i\}$ carry $4$ units of flow, and the third arc does not carry any flow.

        Suppose that $x(u^iy_1^i) = 4$. Hence, we have $x(y_1^iy_2^i) = 4$. Since $\Delta^+(D_x) \leq 2$, and because $4$ units of flow enter $y_2^i$, we must have $x(y_2^iy_3^i)=x(y_2^iy_4^i)=x(y_3^iy_4^i) = 2$ and $x(y_2^ir_j) = 0$ (where $j$ is the index of the second clause $C_j$ containing $\varsymb_i$). Hence $4$ units of flow enter $y_4^i$, and repeating the same arguments we obtain $x(y_7^iv^i) = 4$, which shows $x(u^iy_1^i) = x(y_7^iv^i)$. 

        Similarly, if $x(u^iz_1^i) = 4$, we have $x(z_1^iz_2^i) = x(y_7^iv^i) = 4$, $x(z_2^iz_3^i)=x(z_2^iz_4^i)=x(z_3^iz_4^i) = 2$ and $x(z_2^ir_\ell) = 0$ (where $\ell$ is the index of the second clause $C_j$ containing $\bar{\varsymb}_i$).

        Hence at least $8$ units of flow enter $v^i$. Since $v^i$ has exactly one leaving arc (namely $v^iu^{i+1}$ if $i\leq n-1$ and $v^is$ otherwise) with capacity $8$, there must be exactly $8$ units of entering $v^i$.
        Thus, when $i\leq n-1$, we have $x(v^iu^{i+1}) = 8$, and one of $u^iv^i, y_7^iv^i, z_7^iv^i$ does not carry any flow. 

        It remains to show that $x(u^iv^i) > 0$.
        Assume for a contradiction that $x(u^iv^i) = 0$, that is $x(u^iy_1^i)=x(u^iz_1^i) = 4$. Let $x'$ be the flow defined as follows:
        \[
        x'(a) = \left\{
        \begin{array}{ll}
            4 & \mbox{if } a = u^iv^i  \\
            0 & \mbox{if } a\in \{u^iz_1^i,z_2^iz_3^i,z_2^iz_4^i,z_3^iz_4^i,z_4^iz_5^i,z_5^iz_6^i,z_5^iz_7^i,z_6^iz_7^i,z_7^iv^i\}\\
            x(a) & \mbox{otherwise.}\\
        \end{array}
        \right.
        \]
        By definition, we have $|x'| = |x|$, $\Delta^+(D_x)\leq 2$ and $|A(D_{x'})| = |A(D)| - 8$. Hence $x'$ contradicts the choice of $x$, and the claim follows.
    \end{subproof}

    \begin{claim}
        \label{claim:structure_x_1_side}
        There exists a path-flow $P$ of $x$ that contain all vertices $q_1,r_1,q_2,r_2,\dots,q_m,r_m$ in this order.
    \end{claim}
    \begin{subproof}
        We proceed by induction on $j\in [m]$. Since $|x| = 9$, we necessarily have $x(sq_1) = 1$. Let $P$ be the unique path-flow containing the arc $sq_1$.

        Assume that, for some $j\in [n]$, $P$ contains all vertices $q_1,r_1,\dots,r{j-1},q_j$ in this order. 
        As one unit of flow enters $q_j$, the successor of $q_j$ in $P$ is a vertex $w_j\in V(W^i)$ for some $i\in [n]$. Let $w_j'$ be the successor of $w_j$ in $P$, and observe that $w_j' \in \{z_2^i,z_5^i,y_2^i,y_5^i\}$. Assume that $w_j'\in \{y_2^i,y_5^i\}$ (so $w_j \in \{y_1^i,y_4^i\}$), the other case being symmetric. By Claim~\ref{claim:structure_x_9_side}, we have $x(u^iy_1^i)=x(y_7^iv^i) \in \{0,4\}$. Since both $y_1^i$ and $y_4^i$ have only one leaving arc with capacity $4$, we must have $x(u^iy_1^i)=x(y_7^iv^i)=0$. 
        
        If $w_j' = y_5^i$, then $r_j$ must be the successor of $w_j'$, for otherwise $x(y_7^iv^i)>0$, a contradiction. Else if $w_j' = y_2^i$, we claim that $r_j$ is the successor of $w_j'$ in $P$. Assume for a contradiction that its successor belongs to $\{y_3^i,y_4^i\}$. In both cases, $P$ contains the arc $y_4^iy_5^i$. Since $x(y_7^iv^i)=0$, the successor of $y_5^i$ must be $r_{j'}$ for some $j\in [m]$. By construction, we must have $j'<j$. Hence, by induction hypothesis on $P$, $P$ contains a cycle, a contradiction.

        This shows that $P$ contains $q_1,r_1,q_2,r_2,\dots,q_j,r_j$ in this order. If $j\leq m-1$, $P$ contains $q_1,r_1,q_2,r_2,\dots,q_j,r_j,q_{j+1}$ in this order as $q_{j+1}$ is the only out-neighbour of $r_j$, so the claim follows.~
    \end{subproof}

    We are now ready to prove that $\mathcal{F}$ is satisfiable.
    Let $\phi$ be the truth assignment which is true on variable $\varsymb_i$ if and only if $x(u^iz_1^i)=4$. Let us show that $\mathcal{F}$ is satisfied by $\phi$. Let $C_j\in \mathcal{C}$ be any clause. By Claim~\ref{claim:structure_x_1_side}, the path-flow $P$ contains $q_j$. Let $w_j \in \{z_1^i,z_4^i,y_1^i,y_4^i\}$ (for some $i\in [n]$) be its successor in $P$. If $w_j\in \{y_1^i,y_4^i\}$, by construction $\varsymb_i$ appears positively in $C_j$. By Claim~\ref{claim:structure_x_9_side}, and because both $y_1^i$ and $y_4^i$ have only one leaving arc with capacity $4$, we must have $x(u^iy_1^i) = 0$. Hence, $x(u^iz_1^i)=4$ and $\phi$ is true on $\varsymb_i$, so $C_j$ is satisfied by $\phi$. 
    Else if $w_j\in \{z_1^i,z_4^i\}$, by construction $\varsymb_i$ appears negatively in $C_j$. By Claim~\ref{claim:structure_x_9_side}, and because both $z_1^i$ and $z_4^i$ have only one leaving arc with capacity $4$, we must have $x(u^iz_1^i) = 0$. Hence, $\phi$ is false on $\varsymb_i$, so $C_j$ is satisfied by $\phi$.
    This concludes the proof  that {\sc $(\Delta^+ \le 2)$-flow of value $9$} is NP-hard.
\end{proof}

We easily derive the following inapproximability result.

\begin{corollary}
    \label{cor:inapx_dplus_2}
    Unless P=NP, {\sc $(\Delta^+ \le 2)$-Max-Flow} cannot be approximated by any ratio larger than $\frac{8}{9}$.
\end{corollary}

We note that the proof of Theorem~\ref{thm:npc_dplus} can be adapted to show that the problem {\sc $(\Delta^+\leq k)$-flow of value $2k^2+1$} is NP-complete for every fixed $k\geq 2$. To do so, modify the variable-gadget $W$ as follows:
\begin{enumerate*}[label=(\roman*)]
  \item replace the two paths  from $y_2$ to $y_4$ (respectively $y_5$ to $y_7$, $z_2$ to $z_4$, and $z_5$ to $z_7$) by $k$ paths with capacity $2$,
  \item replace the arc $uv$ by $k-1$ paths each with capacity $2k$, and 
  \item set the capacity of the eight remaining arcs to $2k$.
\end{enumerate*}
Finally, in the network $\mathcal{N}_{\mathcal{F}}$, the arcs with capacity $8$ receive capacity $2k^2$. The same arguments show that $\mathcal{F}$ is a positive instance of $(3,B2)$-SAT if and only if the obtained network admits a $(\Delta^+\leq k)$-flow of value $2k^2+1$.

Hence, for fixed $k$, if $\ell$ is large enough, deciding the existence of a $(\Delta^+\leq k)$-flow of value $k+\ell$ is an NP-complete problem. We indeed believe that $\ell$ being large enough is a necessary condition for this problem to be NP-complete, and pose the following conjecture.

\begin{conjecture}
    For every fixed integer $\ell$, there exists $k_\ell\in \mathbb{N}$ such that, for every $k\geq k_\ell$, {\sc $(\Delta^+\leq k)$-flow of value $k+\ell$} is solvable in polynomial time.
\end{conjecture}

\subsection{Bounded maximum in- and out-degree}

In this section we consider the problems of the two previous sections, but now we ask $D_x$ to have bounded maximum in- and out-degrees (and not only bounded maximum out-degree). We will justify that the complexity results are exactly the same as in the previous sections. We start with the following one.

\defproblem
{\sc $(\Delta^+ \le k_1, \Delta^- \le k_2)$-Max-Flow}{A flow network ${\cal N}=(D,s,t,c)$}
{The maximum value of a flow $x$ on $\cal N$ such that $\Delta^+(D_x)\le k_1$ and $\Delta^-(D_x)\le k_2$.}

\begin{figure}[hbtp]
    \begin{minipage}{\linewidth}
        \begin{center}	
          \begin{tikzpicture}[thick,scale=1, every node/.style={transform shape}]
            \tikzset{vertex/.style = {circle,fill=black,minimum size=4pt, inner sep=0pt}}
    	\tikzset{edge/.style = {->,> = latex'}}
     
            \node[vertex, label=left:$y_1$] (y1) at (0,0) {};
            \node[vertex, label=left:$y_2$] (y2) at (1.5,0) {};
            \node[] (dots1) at (3,0) {$\cdots$};
            \node[vertex, label=left:$y_m$] (y3) at (5,0) {};
            
            \node[vertex, label=below:$t_1$] (t1) at (0,-1) {};
            \node[vertex, label=below:$t_2$] (t2) at (1.5,-1) {};
            \node[] (dots2) at (3,-1) {$\cdots$};
            \node[vertex, label=below:$t_m$] (t3) at (5,-1) {};
            
            \node[vertex, label=below:$t$] (t) at (6.5,-1) {};
            \node[vertex, label=right:$u_{m+1}$] (um) at (6.5,1) {};
            
            \draw[edge] (y1) -- (t1) node [midway,xshift=-0.6em] () {\scriptsize{$1$}};
            \draw[edge] (t1) -- (t2) node [midway,yshift=-0.6em] () {\scriptsize{$1$}};
            
            \draw[edge] (y2) -- (t2) node [midway,xshift=-0.6em] () {\scriptsize{$1$}};
            \draw[edge] (t2) -- (dots2) node [midway,yshift=-0.6em] () {\scriptsize{$2$}};
            
            \draw[edge] (dots1) -- (dots2) node [midway,xshift=-0.6em] () {\scriptsize{$1$}};
            \draw[edge] (dots2) -- (t3) node [midway,yshift=-0.6em] () {\scriptsize{$m-1$}};
            
            \draw[edge] (y3) -- (t3) node [midway,xshift=-0.6em] () {\scriptsize{$1$}};
            \draw[edge] (t3) -- (t) node [midway,yshift=-0.6em] () {\scriptsize{$m$}};
            
            \draw[edge] (um) -- (t) node [midway,xshift=1.4em] () {\scriptsize{$2m+1$}};
            
          \end{tikzpicture}
      \caption{A gadget that ensures $d^-(t) \leq 2$ in the support of every flow.}
      \label{fig:delta-_2}
    \end{center}    
  \end{minipage}
\end{figure}
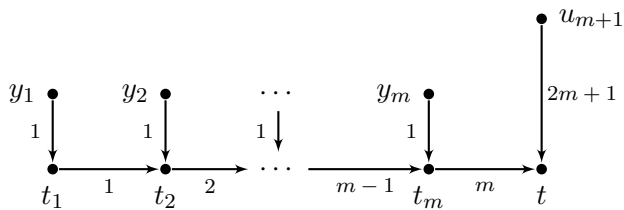

If $k_1$ or $k_2$ is equal to one, then {\sc $(\Delta^+ \le k_1, \Delta^+ \le k_2)$-Max-Flow} consists of finding the maximum $c$ such that there exists a path from $s$ to $t$ in which every arc has capacity at least $c$. This can be done in polynomial time.
Now if $k_1, k_2 \geq 2$, {\sc $(\Delta^+ \le k_1, \Delta^+ \le k_2)$-Max-Flow} is NP-hard even when restricted to acyclic networks. We use the same reduction as in the proof of Theorem~\ref{thm:Delta2acyclicflow}. In the built network, for every flow $x$ with $\Delta^+(D_x) \leq 2$, every vertex but $t$ has in-degree at most two. Let us replace the single vertex $t$ by the in-branching depicted in Figure~\ref{fig:delta-_2}. In the obtained network, for every flow $x$ with $\Delta^+(D_x) \leq 2$, we have $\Delta^-(D_x) \leq 2$. Since $k_2 \geq 2$, this shows the hardness of the problem.

Analogously, one can check that the reduction made in the proof of Theorem~\ref{thm:npc_dplus} shows that deciding whether a network $\mathcal{N}$ admits a $(\Delta^+\leq 2, \Delta^-\leq 2)$-flow of value $9$ is an NP-complete problem.

\section{Flows with high connectivity}\label{sec:highconflow}

In this section we turn our focus on maximum flows $|x|$ for which $\lambda_{D_x}(s,t)$ is large enough. We first prove that a trivial necessary condition on a network $\mathcal{N}=(D,s,t,c)$ to admit a maximum flow $x$ with $\lambda_{D_x}(s,t) \geq 2$ is indeed sufficient.
In a network ${\cal N}=(D,s,t,c)$, an {\bf $(s,t)$-cut-vertex} (respectively an \textbf{$(s,t)$-cut-arc}) is a vertex in $V(D) \setminus \{s,t\}$ (respectively an arc in $A(D)$) which is contained in all the paths of $D$ from $s$ to $t$.

\begin{theorem}
\label{theo:lambda2}
If $\lambda_D(s,t)\ge 2$, then there exists a maximum flow $x$ on $\cal
N$ such that $\lambda_{D_x}(s,t)\ge 2$. In particular, it is
polynomial time solvable to decide if a network $\cal N$ admits a maximum
flow $x$ with $\lambda_{D_x}(s,t)\ge 2$.
\end{theorem}

\begin{proof}
As having $\lambda_D(s,t)\ge 2$ is a necessary condition to get a
maximum flow $x$ on $\cal N$ such that $\lambda_{D_x}(s,t)\ge 2$, and
as it is possible to decide in polynomial time if $\lambda_D(s,t)\ge
2$ or not, the first part of the statement implies the announced polynomial time algorithm.

Let us assume then that $\lambda_D(s,t)\ge 2$
and let us consider a maximum flow $x$ on $\cal N$ such that $D_x$ has a minimum number of  $(s,t)$-cut-arcs. Let us show then that $D_x$ has no $(s,t)$-cut-arc. 

Towards a contradiction, assume that $D_x$ contains some  $(s,t)$-cut-arcs. Consider a path $Q$ from $s$ to $t$ in $D_x$. By definition of an  $(s,t)$-cut-arc, every  $(s,t)$-cut-arc of $D_x$ belongs to $Q$. We label these  $(s,t)$-cut-arcs $u_1v_1,\dots,u_\ell v_\ell$ according to their ordering along $Q$. In what follows, we identify $t$ with $u_{\ell+1}$ and $s$ with $v_0$.

Every vertex of $D_x$ is on a path from $s$ to $t$. As every path from $s$ to $t$ contains all the  $(s,t)$-cut-arcs $u_iv_i$, we can partition the vertices of $D_x$ into $(X_0,\dots,X_\ell)$ where $X_i=\{y\in V(D_x)\ : \ y\textrm{ belongs to a path from }v_i\textrm{ to }u_{i+1}\textrm{ in } D_x\}$ for every $i\in \{0,\dots,\ell\}$. 
For every $i\in \{0,\dots,\ell\}$, note that there exist two arc-disjoint paths $Q_i^1,Q_i^2$ from $v_{i}$ to $u_{i+1}$, for otherwise there exists a $(v_i,u_{i+1})$-cut-arc in $D_x$, which is also an $(s,t)$-cut-arc.

Since $u_1v_1$ is not a  $(s,t)$-cut-arc in $D$, $D$ contains a path $P=y_1,\dots,y_{|P|}$ such that $V(P) \cap X_0 = \{y_1\}$ and $V(P) \cap (X_1\cup \dots X_{\ell}) = \{y_{|P|}\}$. 

Let $j\geq 1$ be the index such that $y_{|P|} \in X_j$.
We assume without loss of generality that $Q_0^1$ contains $y_1$ and $Q_j^1$ contains $y_{|P|}$. Let $Q'$ be the path made of the concatenation of all the $Q_i^1$s, and $Q$ be its subpath from $y_1$ to $y_{|P|}$. We define the $(s,t)$-flow $x'$ of $\mathcal{N}$ as follows:
\[
\forall e\in A(D), x'(e) = \left\{
    \begin{array}{ll}
        1 & \mbox{if } e \in A(P) \\
        x(e) - 1 & \mbox{if } e\in A(Q)\\
        x(e) & \mbox{otherwise.} 
    \end{array}
\right.
\]

Clearly $x'$ is an $(s,t)$-flow and $|x'| = |x|$. Observe that $D_{x'}$ contains two arc-disjoint paths from $s$ to $u_{j+1}$: one is the concatenation $P_1' = Q_0^2u_1v_1Q_1^2u_2v_2,\dots,u_{j}v_jQ_j^2$ and the other is the concatenation $P_2'$ of $Q_0^1[s,y_1]$, $P$ and $Q_j^1[y_{|P|},u_{j+1}]$. Note that $P_2'$ is disjoint from $Q$ so it is actually a path in $D_{x'}$. Also, $P_1'$ intersects $Q$ exactly on the arcs $u_iv_i$ for $i\leq j-1$, and all these arcs carry at least two units of flow since they are  $(s,t)$-cut-arcs (and $\lambda_D(s,t) \geq 2$ implies $|x| \geq 2$).
See Figure~\ref{fig:max_flow_connectivity_2} for an illustration.

    \ifthenelse{\boolean{colouredfigures}}
    {
    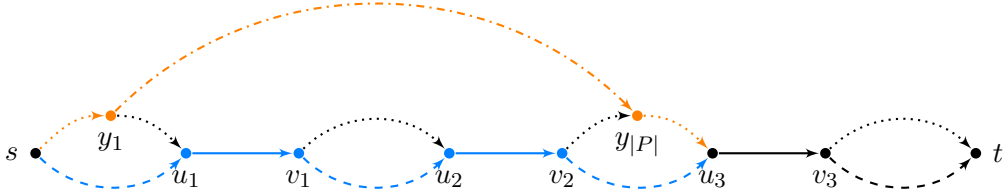
\begin{figure}[hbtp]
        \begin{center}	
              \begin{tikzpicture}[thick,scale=1, every node/.style={transform shape}]
                \tikzset{vertex/.style = {circle,fill=black,minimum size=4pt, inner sep=0pt}}
                \tikzset{littlevertex/.style = {circle,fill=black,minimum size=0pt, inner sep=0pt}}
    	    \tikzset{edge/.style = {->,> = latex'}}
                \tikzset{bigvertex/.style = {shape=circle,draw, minimum size=2em}}

                \node[vertex, label=left:$s$] (s) at (0,0){};
                \node[vertex, orange, label=below:$y_1$] (y1) at (1,0.5){};
                \node[vertex, g-blue, label=below:$u_1$] (w1m) at (2,0) {};
                \node[vertex, g-blue, label=below:$v_1$] (w1p) at (3.5,0) {};
                \node[vertex, g-blue, label=below:$u_2$] (w2m) at (5.5,0) {};
                \node[vertex, g-blue, label=below:$v_2$] (w2p) at (7,0) {};
                \node[vertex, orange, label=below:$y_{|P|}$] (yp) at (8,0.5){};
                \node[vertex, label=below:$u_3$] (w3m) at (9,0) {};
                \node[vertex, label=below:$v_3$] (w3p) at (10.5,0) {};
                \node[vertex, label=right:$t$] (t) at (12.5,0){};
                
                \draw[edge,dotted, orange] (s) to[in=180, out=45] (y1){};
                \draw[edge,dotted] (y1) to[in=135, out=0] (w1m){};
                \draw[edge, g-blue, dashed] (s) to[in=-135, out=-45] (w1m){};
                \draw[edge, g-blue] (w1m) -- (w1p) ;
                
                \draw[edge,dotted] (w1p) to[in=135, out=45] (w2m){};
                \draw[edge, g-blue, dashed] (w1p) to[in=-135, out=-45] (w2m){};
                \draw[edge, g-blue] (w2m) -- (w2p);
                
                \draw[edge,dotted] (w2p) to[in=180, out=45] (yp){};
                \draw[edge,dotted,orange] (yp) to[in=135, out=0] (w3m){};
                \draw[edge, g-blue, dashed] (w2p) to[in=-135, out=-45] (w3m){};
                \draw[edge] (w3m) -- (w3p);
                
                \draw[edge,dotted] (w3p) to[in=135, out=45] (t){};
                \draw[edge,dashed] (w3p) to[in=-135, out=-45] (t){};
                
                \draw[edge, dash dot, orange] (y1) to[in=135, out=45] (yp){};
              \end{tikzpicture}

          \caption{The support $D_x$ and the construction of $x'$. On this example, we have $\ell=3$ and $j=2$. The solid arcs represent the $(s,t)$-cut-arcs. The dotted and dashed arcs represent the paths $Q_i^1$ and $Q_i^2$ respectively. The path $P$ is represented by dash-dotted lines. The paths $P_1'$ and $P_2'$ are represented respectively in blue and orange.}
          
          \label{fig:max_flow_connectivity_2}
        \end{center}
    \end{figure}
    }
    {
    \begin{figure}[hbtp]
        \begin{center}	
              \begin{tikzpicture}[thick,scale=1, every node/.style={transform shape}]
                \tikzset{vertex/.style = {circle,fill=black,minimum size=4pt, inner sep=0pt}}
                \tikzset{littlevertex/.style = {circle,fill=black,minimum size=0pt, inner sep=0pt}}
    	    \tikzset{edge/.style = {->,> = latex'}}
                \tikzset{bigvertex/.style = {shape=circle,draw, minimum size=2em}}

                \node[vertex, label=left:$s$] (s) at (0,0){};
                \node[vertex, label=below:$y_1$] (y1) at (1,0.5){};
                \node[vertex, label=below:$u_1$] (w1m) at (2,0) {};
                \node[vertex, label=below:$v_1$] (w1p) at (3.5,0) {};
                \node[vertex, label=below:$u_2$] (w2m) at (5.5,0) {};
                \node[vertex, label=below:$v_2$] (w2p) at (7,0) {};
                \node[vertex, label=below:$y_{|P|}$] (yp) at (8,0.5){};
                \node[vertex, label=below:$u_3$] (w3m) at (9,0) {};
                \node[vertex, label=below:$v_3$] (w3p) at (10.5,0) {};
                \node[vertex, label=right:$t$] (t) at (12.5,0){};
                
                \draw[edge,dotted] (s) to[in=180, out=45] (y1){};
                \draw[edge,dotted] (y1) to[in=135, out=0] (w1m){};
                \draw[edge, dashed] (s) to[in=-135, out=-45] (w1m){};
                \draw[edge] (w1m) -- (w1p) ;
                
                \draw[edge,dotted] (w1p) to[in=135, out=45] (w2m){};
                \draw[edge, dashed] (w1p) to[in=-135, out=-45] (w2m){};
                \draw[edge] (w2m) -- (w2p);
                
                \draw[edge,dotted] (w2p) to[in=180, out=45] (yp){};
                \draw[edge,dotted] (yp) to[in=135, out=0] (w3m){};
                \draw[edge, dashed] (w2p) to[in=-135, out=-45] (w3m){};
                \draw[edge] (w3m) -- (w3p);
                
                \draw[edge,dotted] (w3p) to[in=135, out=45] (t){};
                \draw[edge,dashed] (w3p) to[in=-135, out=-45] (t){};
                
                \draw[edge, dash dot] (y1) to[in=135, out=45] (yp){};
            \end{tikzpicture}

          \caption{The support $D_x$ and the construction of $x'$. On this example, we have $\ell=3$ and $j=2$. The solid arcs represent the $(s,t)$-cut-arcs. The dotted and dashed arcs represent the paths $Q_i^1$ and $Q_i^2$ respectively. The path $P$ is represented by dash-dotted lines.}
          
          \label{fig:max_flow_connectivity_2}
        \end{center}
    \end{figure}
    }
    Hence every  $(s,t)$-cut-arc of $D_{x'}$ must be on every path from $u_{j+1}$ to $t$, implying that it is also a  $(s,t)$-cut-arc in $D_x$. Since $u_1v_1$ is not a  $(s,t)$-cut-arc in $D_{x'}$, $x'$ contradicts the choice of $x$.
\end{proof}

In the following we show that it is not possible to generalize the first part of
Theorem~\ref{theo:lambda2} for higher values of $\lambda_D(s,t)$.

\begin{theorem}
    For every fixed integer $\lambda \geq 3$, there exists a network $\mathcal{N}_k=(D,s,t,c)$ such that $\lambda_D(s,t) = \lambda$ and for every maximum $(s,t)$-flow $x$, $\lambda_{D_x}(s,t) \leq 2$.
\end{theorem}

\begin{proof}
    Let $D$ be the digraph made of $\lambda-2$ $(s,t)$-paths of length $3$ $su_iv_it$, $i\in [\lambda-2]$ and two $(s,t)$-paths of length $2$ $syt$ and $szt$.
    Then add all the arcs of $\{u_iy \mid i\in [\lambda-2]\} \cup \{zv_i \mid i\in [\lambda-2]\}$. All the arcs have capacity 1 except $sz$ and $yt$ which have capacity $\lambda-1$. See Figure~\ref{fig:lambdak} for an illustration.
    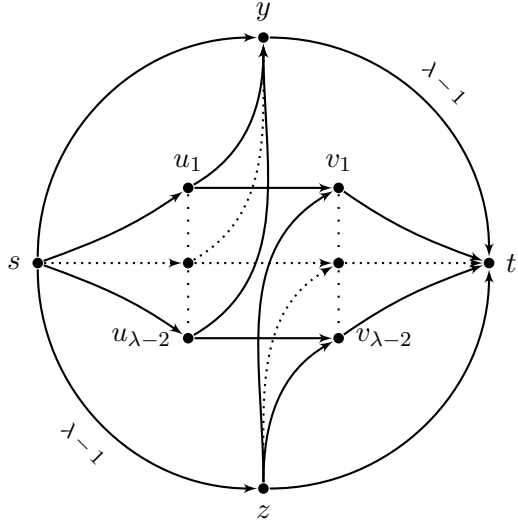
\begin{figure}[hbtp]
        \begin{center}	
              \begin{tikzpicture}[thick,scale=1, every node/.style={transform shape}]
                \tikzset{vertex/.style = {circle,fill=black,minimum size=4pt, inner sep=0pt}}
                \tikzset{littlevertex/.style = {circle,fill=black,minimum size=0pt, inner sep=0pt}}
    	    \tikzset{edge/.style = {->,> = latex'}}
                \tikzset{bigvertex/.style = {shape=circle,draw, minimum size=2em}}

                \node[vertex, label=left:$s$] (s) at (-1,0){};
                \node[vertex, label=right:$t$] (t) at (5,0){};
                \node[vertex, label=above:$y$] (y) at (2,3){};
                \node[vertex, label=below:$z$] (z) at (2,-3){};
                
                \node[vertex, label=above:$u_1$] (u1) at (1,1){};
                \node[vertex] (u2) at (1,0){};
                \node[vertex, label=left:$u_{\lambda-2}$] (u3) at (1,-1){};
                
                \node[vertex, label=above:$v_1$] (v1) at (3,1){};
                \node[vertex] (v2) at (3,0){};
                \node[vertex, label=right:$v_{\lambda-2}$] (v3) at (3,-1){};
                
                \draw[edge] (u1) to (v1);
                \draw[edge,dotted] (u2) to (v2);
                \draw[edge] (u3) to (v3);
                \draw[loosely dotted] (u1) -- (u3){};
                \draw[loosely dotted] (v1) -- (v3){};

                \draw[edge] (u1) to[out=30, in=-90] (y);
                \draw[edge,dotted] (u2) to[out=30, in=-90] (y);
                \draw[edge] (u3) to[out=30, in=-90] (y);

                \draw[edge] (z) to[out=90, in=-150] (v1);
                \draw[edge,dotted] (z) to[out=90, in=-150] (v2);
                \draw[edge] (z) to[out=90, in=-150] (v3);

                \draw[edge] (s) to[out=20, in=-145] (u1) {};
                \draw[edge,dotted] (s) to (u2) {};
                \draw[edge] (s) to[out=-20, in=145] (u3) {};
                
                \draw[edge] (v1) to[out=-35, in=160] (t) {};
                \draw[edge,dotted] (v2) to (t) {};
                \draw[edge] (v3) to[out=35, in=-160] (t) {};

                \draw[edge] (s) to[out=-90, in=180] (z){};
                \draw[edge] (s) to[out=90, in=180] (y){};
                
                \draw[edge] (y) to[out=0, in=90] (t){};
                \draw[edge] (z) to[out=0, in=-90] (t){};

                \node[rotate=-45] (c1) at (-0.4,-2.4) {\scriptsize $\lambda-1$};
                \node[rotate=-45] (c1) at (4.4,2.4) {\scriptsize $\lambda-1$};
                
              \end{tikzpicture}
            \caption{Example of a network (all arcs have capacity 1 except the two with indicated capacities $\lambda-1$). The network has $\lambda$ arc-disjoint paths from $s$ to $t$, but  for every maximum flow $x$ (there is only one) there exists only  two arc-disjoint paths from $s$ to $t$ in $D_x$.}
          \label{fig:lambdak}
        \end{center}
    \end{figure}

    We clearly have $\lambda_D(s,t) = \lambda$, let us show that for every maximum $(s,t)$-flow $x$, $\lambda_{D_x}(s,t) = 2$. 
    Observe first that there exists a flow $x^*$ of value $2\lambda-2$, which is defined as follows:
    \[
        x^*(uv) = \left\{
        \begin{array}{ll}
            \lambda-1 & \mbox{if } uv\in \{sz, yt\}  \\
            0 & \mbox{if } uv\in \{u_iv_i \mid i\in [\lambda-2]\}  \\
            1 & \mbox{otherwise.}\\
        \end{array}
    \right.
    \]
    
    Let $x$ be a maximum $(s,t)$-flow, then we claim that $D_x$ does not contain any arc of the form $u_iv_i$, for $i\in [\lambda-2]$. Assume for a contradiction that it does. Let $P$ be the path-flow of $x$ containing the arc $u_iv_i$. Since $u_i$ has in-degree 1 and $su_i$ has capacity 1, we have $x(u_iy) = 0$. Since $y$ has in-degree $\lambda-1$, we have $x(yt) \leq \lambda-2$. Hence the flow entering $t$ is at most $|x| \leq (\lambda-1)+(\lambda-2) = 2\lambda-3$, a contradiction to the maximality of $x$ since $|x^*| = 2\lambda-2 > |x|$.

    This shows that $D_x$ does not contain any arc of the form $u_iv_i$. Hence $D_x \setminus \{sz,yt\}$ does not contain any $(s,t)$-path, implying that $\lambda_{D_x}(s,t) \leq 2$ as desired.
\end{proof}

\section{Safe maximum flows}\label{sec:fragile}

We now discuss problems concerning stability of $(s,t)$-flows in a given  network towards arc-deletions. Related problems for other structures were considered in~\cite{bangTCS526,bangDAM209,bangTCS595,bessyArxiv2105.01582,DISSER202018}.

If we delete a set $A'$ of one or more arcs from the support $D_x$ of a maximum  $(s,t)$-flow $x$, then the restriction of $x$ to $A(D_x)\setminus A'$ may not be an $(s,t)$-flow anymore, but it can be modified to a new maximum flow $\tilde{x}$ in $D_x[A(D_x)\setminus A']$ of value at least $|x|-\sum_{uv\in A'}x(uv)$ in polynomial time.

Consider finding in a fixed network $\cal N$ a maximum flow $x^*$ which is as persistent as possible against (a few) arcs deletions. That is, the value of $x^*$ restricted to the arc set $A\setminus A'$, where $|A'|\leq k$ is as high as possible. 
Clearly, if we delete one or more arcs that are in the same   minimum $(s,t)$-cut of $\cal N$, then every maximum flow will be hurt by the same amount (the sum of the capacities of the deleted arcs) but this is by no means the case for arcs which are not contained in some min cut: one max flow may send all its flow through some arc of very high capacity (not in any min cut)  and hence be hurt very much if that arc is deleted, while another max flow may send its flow along many pairwise arc-disjoint paths.

For a given flow network ${\cal N}=(D=(V,A),s,t,c)$ we can determine in polynomial time the subset $A_{mincut}\subseteq A$ of arcs that belong to some minimum $(s,t)$-cut in $\cal N$.

\defproblem
{\sc $k$-Arc-Persistent-Max-Flow}
{A flow network ${\cal N}=(D=(V,A),s,t,c)$ and an integer $k$.}
{A maximum flow $x$ such that the minimum value of a maximum flow $\tilde{x}$ in $D_x[A(D_x)\setminus A']$ over all subsets $A'\subseteq A\setminus A_{mincut}$ of $k$ arcs is as large as possible.}

This problem turns out to be NP-hard even when $k=1$ and all capacities are 1. Indeed, let $\mathcal{N}=(D,s,t,c)$ be a network and assume that $s$ has exactly one leaving arc $ss'$ and $t$ has exactly one entering arc $t't$, so the maximum flow is $1$. Assume also that there exists two arc-disjoint paths from $s'$ to $t'$, so $A_{mincut}$ is exactly $\{ss',t't\}$. Then the problem consists of deciding whether $\mathcal{N}$ admits an $(s,t)$-flow $x$ such that, for every arc $uv\notin A_{mincut}$, $D_x$ contains an $(s,t)$-path disjoint from $uv$. We claim that this problem consists exactly of finding, in $D - \{s,t\}$, three arc-disjoint paths $P_1,P_2$ and $P_3$ such that $P_1,P_2$ go from $s'$ to $t'$ and $P_3$ goes from $t'$ to $s'$, which is known to be NP-hard. This follows from results in~\cite{fortuneTCS10} and the fact that we can reduce a problem on vertex-disjoint paths to a problem on arc-disjoint paths via the vertex splitting operation. 

To see this, assume first that we have such three paths $P_1,P_2$ and $P_3$. Let $x$ be the flow obtained by sending exactly one unit of flow along each path and one unit of flow along $ss'$ and $tt'$. Then, for every arc $uv\notin A_{mincut} = \{ss',t't\}$, $D_x \setminus \{uv\}$ contains at least one of the paths $sP_1t$ and $sP_2t$.

Conversely, if $\mathcal{N}$ admits such a flow $x$, then it can be decomposed into  a path-flow $sP_1t$ (where $P_1$ goes from $s'$ to $t'$) and some cycle-flows $C_1,\dots,C_r$. Note that $P_1,C_1,\dots,C_r$ are pairwise arc-disjoint because all the capacities are 1. The digraph $D'$ on vertices $V(D) \setminus \{s,t\}$ with arc-set $A(C_1)\cup \dots \cup A(C_r)$ is eulerian, and $s,t$ must belong to the same  connected component of $D'$, for otherwise the last arc of $P_1$ leaving the connected component of $s$ is an $(s,t)$-cut-arc in $D_x$. Hence, since $D'$ is eulerian, it admits two arc-disjoint paths $P_2,P_3$ such that $P_2$ goes from $s'$ to $t'$ and $P_3$ goes from $t'$ to $s'$.

For larger values of $k$, one can just add $k-1$ arcs (of paths of length two if we forbid multiple arcs) from $s$ to $t$ to obtain the hardness result.

The perspicacious reader may see that, in the proof above, the hardness of the {\sc $k$-Arc-Persistent-Max-Flow} problem mainly comes from the fact that an optimal solution may contain cycle-flows. The following problem then naturally arises.

\begin{problem}
    What is the complexity of the {\sc 1-Arc-Persistent-Max-Flow} problem for acyclic digraphs?
\end{problem}

There is an analogous version involving vertex deletions. Again, if a vertex $v$ is incident to one or more arcs across some minimum $(s,t)$-cut, then deleting $v$ and its incident arcs will affect the value of all such  minimum $(s,t)$-cuts by the same amount. Let $V_{mincut}$ be the set of vertices of $\cal N$ which are incident to at least one arc in a minimum $(s,t)$-cut.

\defproblem
{\sc $k$-Vertex-Persistent-Max-Flow}
{A flow network ${\cal N}=(D=(V,A),s,t,c)$ and an integer $k$.}
{A maximum flow $x$ such that the minimum value of a maximum flow $\tilde{x}$ in $D_x[V(D_x)-V']$ over all subsets $V'\subseteq V-V_{mincut}$ of $k$ vertices is as large as possible.}

Using exactly the same arguments as above, one can show that already {\sc 1-vertex persistent-Max-Flow problem} is NP-hard with all capacities are $1$, by reducing from the vertex-disjoint version of the problem of finding three paths $P_1$, $P_2$, and $P_3$ such that $P_1,P_2$ go from $s'$ to $t'$ and $P_3$ goes from $t'$ to $s'$, which is known to be NP-hard~\cite{fortuneTCS10}.

The following might be an interesting generalisation of arc-connectivity.

\begin{question}
    What is the complexity of finding, for a given network ${\cal N}$ and an integer $K$, the minimum number of arcs that one has to delete before the max-flow value in the resulting network is below $K$?
\end{question}

Observe that, when $K=1$, this problem is exactly the min-cut problem. This is why it is actually a generalisation of the arc-connectivity. Note also that, if all the capacities are 1, all the minimum $(s,t)$-cut have the same number of arcs, so the solution is exactly $c(S^*,\Bar{S}) - K +1$ (where $(S^*,\Bar{S^*})$ is any min-cut). 

\section{Flows that are decomposable into a given number of paths}\label{sec:decompintopaths}

\subsection{General case}

In this section we consider the following problems. The first one has been studied previously under the name of the {\bf $p$-splittable} maximum flow problem in~\cite{baierA42}.

\defproblem
{\sc $p$-Decomposable-Max-Flow}
{A flow network ${\cal N}=(D,s,t,c)$ and an integer $p$.}
{The maximum value of a flow $x$ such that $x$ can be decomposed into at most $p$ path-flows.}

\defproblem
{\sc $p$-Arc-Decomposable-Max-Flow}
{A flow network ${\cal N}=(D,s,t,c)$ and an integer $p$.}
{The maximum value of a flow $x$ such that $x$ can be decomposed into at most $p$ pairwise arc-disjoint path-flows.}

\defproblem
{\sc $p$-Vertex-Decomposable-Max-Flow}
{A flow network ${\cal N}=(D,s,t,c)$ and an integer $p$.}
{The maximum value of a flow $x$ such that $x$ can be decomposed into at most $p$ pairwise vertex-disjoint path-flows.}

Note that every solution for the second problem is actually a solution for the first one, but that the converse is not true. Analogously, every solution to the third one is a solution to the second one. 

It was proved in~\cite{baierA42} that {\sc $2$-Decomposable-Max-Flow} is NP-Hard and that it cannot be approximated by any ratio larger than $\frac{2}{3}$ unless P=NP. We extend their result by showing NP-hardness and non-approximability results of {\sc $p$-Decomposable-Max-Flow} for every fixed $p\geq 2$.

\begin{theorem}
    \label{thm:p_decomposable_NPC}
    For any fixed $p\geq 2$, the {\sc $p$-Decomposable-Max-Flow} problem is NP-hard. Moreover, unless P=NP, it cannot be approximated by any ratio larger than $\rho(p) = \min (\rho_1(p), \rho_2(p))$, where $\rho_1(p),\rho_2(p)$ are defined as follows:
    \[
    \rho_1(p) = \left\{
        \begin{array}{ll}
            \frac{5}{6} & \mbox{if } p = 0 \mod 4 \\
            \frac{5p-1}{6p-2} & \mbox{if } p = 1 \mod 4 \\
            \frac{5p-2}{6p} & \mbox{if } p = 2 \mod 4\\
            \frac{5p-3}{6p-2} & \mbox{if } p = 3 \mod 4 \\
        \end{array}
    \right.
    \]
    \[
    \rho_2(p) = \left\{
        \begin{array}{ll}
            \frac{4}{5} & \mbox{if } p \mbox{ is even}\\
            \frac{4p-2}{5p-3} & \mbox{otherwise.}\\
        \end{array}
    \right.
    \]
    
    In particular, $\rho(2) = \frac{2}{3}$, $\rho(3) = \frac{3}{4}$, $\rho(p) \xrightarrow[p \to +\infty ]{} \frac{4}{5}$, and $\rho(p) \leq \frac{9}{11}$ in general.
\end{theorem}

\begin{proof}
    Let us fix $p$, we will first prove that the {\sc $p$-Decomposable-Max-Flow} problem cannot be approximated by any ratio larger than $\rho_1(p)$, and then that it cannot by any ratio larger than $\rho_2(p)$.

     In both cases, we will reduce from the {\sc Weak-$2$-Linkage} problem, which is NP-hard by Theorem~\ref{thm:k_linkage}. Let $\mathcal{I}=(D,s_1,s_2,t_1,t_2)$ be an instance of the {\sc Weak-$2$-Linkage} problem. We assume that there exists an $(s_2,t_2)$-path in $D$, for otherwise $\mathcal{I}$ is clearly a negative instance (and it can be checked in polynomial time). We also assume that there exists two arc-disjoint paths from $\{s_1,s_2\}$ to $\{t_1,t_2\}$, for otherwise $\mathcal{I}$ is clearly a negative instance. 

    In both case, we will build a network $\mathcal{N}=(D',s,t,c)$ from $\mathcal{I}$ and show that the solution of {\sc $p$-Decomposable-Max-Flow} on $\mathcal{N}$ is at least $o^+$ if $\mathcal{I}$ is a positive instance, and at most $o^-$ otherwise, where $o^-<o^+$ are two specific values depending only on $p$. This will show that {\sc $p$-Decomposable-Max-Flow} is NP-Hard and that it cannot be approximated by any ratio larger than $\frac{o^-}{o^+}$ (unless P=NP).

    We now make a construction for which $\frac{o^-}{o^+} = \rho_1(p)$.
    We will distinguish four cases depending on the value of $p$ modulo $4$. They are all similar but we detail each of them for completeness. 

    \begin{itemize}
        \item Assume first that $p = 4q$ for some $q\in \mathbb{N}$.

        We build $\mathcal{N}_0$ as follows. We start from $2q$ disjoint copies $D^1,\dots,D^{2q}$ of $D$. For every vertex $v$ of $D$, we denote its copy in $D^i$ by $v^i$. Then we add a source $s$, a sink $t$, and the arcs $ss_1^{i},ss_2^{i},t_1^it,t_2^it$ for every $i\in [2q]$. All the arcs of $\mathcal{N}_0$ have capacity $2$, except the arcs $ss_1^{i}$ and $t_1^{i}t$ (for $i\in [2q]$) which have capacity $1$.

        \medskip

        Assume first that $\mathcal{I}$ is a positive instance of {\sc Weak-$2$-Linkage}, which means that $D$ contains two arc-disjoint paths $P_1,P_2$ going respectively from $s_1,s_2$ to $t_1,t_2$. For each $i\in [2q]$, let $P_1^i$ and $P_2^i$ be the copies of $P_1$ and $P_2$ respectively in $D^i$. Let $Q_1^i$ and $Q_2^i$ be the concatenations $ss_1^i,P_1^i,t_1^it$ and $ss_2^i,P_2^i,t_2^it$ respectively.
        The flow consisting of exactly two units of flow on every path $Q_2^i$ and exactly one unit of flow on every path $Q_1^i$ shows that the solution of {\sc $p$-Decomposable-Max-Flow} on $\mathcal{N}_0$ is at least $o^+ = 6q$.

        \medskip

        Assume now that $\mathcal{I}$ is a negative instance of {\sc Weak-$2$-Linkage}, and let $x$ be a flow of maximum value that can be decomposed into at most $p$ path-flows. For every $i\in [2q]$, let $x_i = x(ss_1^i) + x(ss_2^i)$. Note that $x_i\leq 3$, we will first show that the maximality of $x$ implies $x_i \geq 2$. 
        
        If $x_i = 1$, then $x$ contains exactly one path-flow going through $D^i$, and it has value $1$. Since $D$ contains a path from $s_2$ to $t_2$, we can replace this path-flow by a path-flow of value 2 to contradict the maximality of $x$. If $x_i=0$, $x$ decomposes into exactly $p$ path-flows, for otherwise we can just add one path-flow to $x$ going through $D^i$ to contradict the maximality of $x$. Then, there exists $j\neq i$ such that $x$ contains three distinct path-flows of value exactly $1$ going through $D^j$. Replacing one of these path-flows by a path-flow of value $2$ going through $D^i$ contradicts the maximality of $x$.

        Let $n$ be the number of indices $i$ such that $x_i = 3$. The number of path-flows $x$ is decomposed into is at least $3n + (2q-n)$. Since this number is at most $p=4q$, it implies that $n\leq q$. Note also that $|x| = 3n + 2(2q-n) = 4q+n$. Hence the value of $x$ is at most $o^- = 5q$.

        We have $\frac{o^-}{o^+} = \frac{5q}{6q} = \frac{5}{6}$ as desired.

        \item Assume now that $p = 4q+1$ for some $q\in \mathbb{N}$.

        We let $\mathcal{N}_1$ be the network obtained from $\mathcal{N}_0$ by adding the arc $st$ with capacity $1$. If $\mathcal{I}$ is a positive instance, using the same flow as in the previous case plus one path-flow of value 1 through the arc $st$, we get that the solution of {\sc $p$-Decomposable-Max-Flow} on $\mathcal{N}_1$ is at least $o^+ = 6q+1$.
        
        \medskip

        Assume now that $\mathcal{I}$ is a negative instance, and let $x$ be a flow of maximum value that can be decomposed into at most $p$ path-flows $P_1,\dots,P_{p'}$. At least one the $P_i$s has value $1$, otherwise for every $i\in [2q]$, there exists exactly one path-flow (which has value $2$) going through $D^i$. Hence $p' = 2q < p$ and we can add to $x$ the path-flow of value $1$ going through the arc $st$, which contradicts the maximality of $x$.

        Hence we may assume without loss of generality that $P_{p'}$ is a path-flow of value $1$, and that $P_{p'}$ is $st$ if one of the $P_i$s is exactly the path on two vertices $st$. Therefore the flow $x'$ made of the $p'-1$ path-flows $P_1,\dots,P_{p'-1}$ is an $(s,t)$ flow of $\mathcal{N}_0$ that can be decomposed into at most $p-1$ path-flows. The previous case implies that $|x'| \leq 5q$. Hence $|x| = |x'| +1 \leq 5q+1 = o^-$.

        We have $\frac{o^-}{o^+} = \frac{5q+1}{6q+1} = \frac{5p-1}{6p-2}$ as desired.

        \item Assume now that $p = 4q+2$ for some $q\in \mathbb{N}$.

        We build $\mathcal{N}_2$ exactly as we build $\mathcal{N}_0$, except that we take $2q+1$ copies of $D$ instead of $2q$. 

        If $\mathcal{I}$ is a positive instance, using exactly the same argument as we did for $\mathcal{N}_0$, we can build a flow on $\mathcal{N}_2$ that can be decomposed into $p$ path-flows and that has value $o^+ = 3(2q+1) = 6q+3$.

        Assume now that $\mathcal{I}$ is a negative instance, and let $x$ be a flow of maximum value that can be decomposed into at most $p$ path-flows. For every $i\in [2q+1]$, let $x_i = x(ss_1^i) + x(ss_2^i)$. For the same reasons as before, we have $x_i \in \{2,3\}$.
        Let $n$ be the number of indices $i$ such that $x_i = 3$. The number of path-flows $x$ decomposed into is at least $3n + (2q+1-n)$, implying that $n\leq q$. Note also that $|x| = 3n + 2(2q+1-n) = 4q+n+2$. Hence the value of $x$ is at most $o^- = 5q+2$.

        We have $\rho(p) = \frac{o^-}{o^+} = \frac{5q+2}{6q+3} = \frac{5p-2}{6p}$ as desired.

        \item Assume finally that $p = 4q+3$ for some $q\in \mathbb{N}$.

        We let $\mathcal{N}_3$ be the network obtained from $\mathcal{N}_2$ by adding the arc $st$ with capacity $1$. Using the same ideas we used in the second case, with $\mathcal{N}_2$ playing the role of $\mathcal{N}_0$, we obtain $o^+ = 6q+4$ and $o^- = 5q+3$. 
        
        Hence we obtain $\frac{o^-}{o^+} = \frac{5q+3}{6q+4} = \frac{5p-3}{6p-2}$ as desired.
        
    \end{itemize}

    This concludes the first part of the proof. We now make a construction for which $\frac{o^-}{o^+} = \rho_2(p)$. Assume first that $p=2q$ is even.
    We build $\mathcal{N}'=(D,s,t,c')$ exactly as $\mathcal{N}_0$ from $q$ copies $D^1,\dots, D^q$ of $D$. We just change the capacities, so every arc has capacity 3 except $ss_1^{i}$ and $t_1^{i}t$ (for $i\in [q]$) which have capacity $2$.

    If $\mathcal{I}$ is a positive instance, then we can send 5 units of flow through each copy of $D$, using exactly $p$ path-flows, and we get a flow of value $5q$. Assume below that $\mathcal{I}$ is a negative instance and let $x$ be a maximum flow in $\mathcal{N}'$. 

    For every $i\in [q]$, let $x_i = x(ss_1^i) + x(ss_2^i)$. The maximality of $x$ implies that $x_i\in \{3, 4, 5\}$ for every $i\in [q]$. Indeed, if all the path-flows of $x$ have value 3, then $x$ decomposed into at most $q-1$ path-flows, and we can just add one additional path-flow of value $3$ through $D^i$. Else if $x$ sends no flow through a copy $D^i$ of $D$, then we can replace a path-flow of $x$ with value $1$ or $2$ by a path-flow of value $3$ through $D^i$.  In both cases, we contradict the maximality of $x$. 
    
    Now, if $x$ has exactly one path-flow through $D^i$, then it has value $3$ (since there exists a path from $s_2$ to $t_2$). If $x$ has two path-flows through $D^i$, then $x_i = 4$ because $\mathcal{I}$ is a negative instance. Moreover, if $x_i = 5$ then $x$ must have $3$ path-flows through $D^i$ (since $\mathcal{I}$ is a negative instance).

    For each $j\in \{3,4,5\}$, let $n_j$ be the number of indices $i$ such that $x_i = j$. We have $|x| = 3\cdot n_3 + 4 \cdot n_4 + 5\cdot n_5$, and the minimum number of path-flows $x$ decomposes into is $n_3 + 2\cdot n_4 + 3\cdot n_5$. If $n_5 > n_3$ then $x$ decomposes into at least $2q+1$ path-flows, a contradiction since $p=2q$. Then we have $n_3\geq n_5$ and $|x| \leq 4\cdot (n_3 + n_4 + n_5) = 4q$. Hence when $p$ is even, unless P=NP, there is no approximation with a ratio larger than $\rho_2(p) = \frac{4q}{5q} = \frac{4}{5}$.
    
    When $p=2q+1$ is odd, we just add to $\mathcal{N}'$ an arc from $s$ to $t$ with capacity $1$. If $\mathcal{I}$ is a positive instance, we have a flow of value $5q+1$. When $\mathcal{I}$ is a negative instance, one can easily show that there exists a maximum flow $x$ that contains the path-flow $st$ with value $1$. Hence, forgetting this path-flow, we obtain a flow on $\mathcal{N}'$, and the result above implies $|x|\leq 4q+1$. Hence, unless P=NP, there is no approximation with a ratio larger than $\rho_2(p) = \frac{4q+1}{5q+1} = \frac{4p-2}{5p-3}$.
\end{proof}

\begin{theorem}
\label{thm:2/3bound}
    For any fixed $p$, both problems {\sc $p$-Arc-Decomposable-Max-Flow} and {\sc $p$-Vertex-Decomposable-Max-Flow} are NP-hard and cannot be approximated by any ratio $\rho > \frac{2}{3}$ unless P=NP.
\end{theorem}
\begin{proof}
    We reduce from {\sc Weak-$2$-Linkage} as in the proof of Theorem~\ref{thm:p_decomposable_NPC} for $p=2$. Since the out-degree of $s$ is 2, any solution to the {\sc $p$-Arc-Decomposable-Max-Flow} problem is actually a solution to the {\sc $2$-Arc-Decomposable-Max-Flow}. This shows the result for the arc-disjoint version.

    The arguments for the vertex-disjoint version are the same, except that we reduce from {\sc 2-Linkage} instead of {\sc 2-Weak-Linkage}.
\end{proof}

It was shown in~\cite{baierA42} that one can obtain a $\rho$-approximation for the value of maximum $p$-decomposable flow in polynomial time, where $\rho = \frac{2}{3}$ if $p\in \{2,3\}$ and $\rho = \frac{1}{2}$ when $p\geq 4$. It is important to note that, in their work, the flows are valued in $\mathbb{R}$, while ours are valued in $\mathbb{N}$.
Still their result can be adapted to flows valued in $\mathbb{N}$, using the following rounding strategy, to obtain a $\frac{\rho}{2}$-approximation. 

First compute in polynomial time a flow $y$ (valued in $\mathbb{R}$) decomposable into $r\leq p$ path-flows $P_1,\dots,P_r$, such that $|y| \geq \rho \cdot |y^*|$, where $y^*$ is a maximum $p$-decomposable flow valued in $\mathbb{R}$. A fortiori, we have $|y| \geq \rho \cdot |x^*|$, where $x^*$ is a maximum $p$-decomposable flow valued in $\mathbb{N}$. Now let $x$ be the $p$-decomposable flow valued in $\mathbb{N}$ with path-flows $P_1',\dots,P_r'$ where, for each $i\in \{1,\dots,r\}$, $P_i$ and $P_i'$ use the same path and $|P_i'| = \left\lfloor |P_i| \right\rfloor$. We thus obtain $|x| \geq \rho\cdot |x^*| - p$. If $|x| < p$, we replace $x$ by any flow of value $p$ which decomposes into $p$ path-flows (we assume that such a flow exists, for otherwise the problem is clearly solvable in polynomial time). We thus have $|x| \geq \max(p,\rho\cdot |x^*| - p)$. If $|x^*| \geq \frac{2p}{\rho}$, then $|x| \geq \rho\cdot |x^*| - p \geq \rho \geq \frac{\rho}{2}|x^*|$. Else if $|x^*| \leq \frac{2p}{\rho}$, then $|x| \geq p \geq \frac{\rho}{2}|x^*|$. 

\medskip

The following result improves on the $\frac{\rho}{2}$-bound for small values of $p$. We will also show that it can be derived into an approximation result for the disjoint versions of the problem.

\begin{theorem}
    \label{thm:apx_decomposable}
    For any fixed $p$, the {\sc $p$-Decomposable-Max-Flow} problem can be approximated by a ratio $\rho = \frac{1}{H(p)}$ where $H(p) = \sum_{i=1}^p\frac{1}{i}$ is the $p^{th}$ term of the harmonic series.
\end{theorem}

\begin{proof}
    Let us fix $p$, we will show that Algorithm~\ref{algo:apx} below is a $\frac{1}{H(p)}$-approximation for the {\sc $p$-Decomposable-Max-Flow} problem. Note first that Algorithm~\ref{algo:apx} runs in polynomial time. In particular, observe that the for loop of line~3 may be replaced by a dichotomy strategy, so if $c_{\max}$ is larger than the size of the network this is not a problem.
    
    \begin{algorithm}
    \textbf{Input}: A flow network ${\cal N}=(D=(V,A),s,t,c)$.
    
    \textbf{Output}: A flow $x$ that is $p$-decomposable  with value at least $\frac{1}{H(p)} \cdot |x^*|$, where $|x^*|$ is the value of an optimal solution.
    
    \begin{algorithmic}[1]
    \For{every $i \in \{1,\dots,p\}$}
        \State $x_i \gets $ the empty flow.
        \For{$\nu = 1$ to $c_{\max}$ where $c_{\max} = \max \{c(uv) \mid uv\in A\}$}
            \State Build $D_\nu$ from $D$ as follows : for each arc $uv$, if $c(uv) < \nu$, then remove $uv$, otherwise replace $uv$ by $\lfloor \frac{c(uv)}{\nu} \rfloor$ multiple arcs.
            \label{line:build_Dv}
            \If{there exists $i$ arc-disjoint paths from $s$ to $t$ in $D_\nu$}\label{line:if_arc_disjoint}
                \State $x_i \gets $ the flow made by sending exactly $\nu$ units of flow on every such path.
            \EndIf
        \EndFor
    \EndFor
    
    \State \Return A flow $x_i$ among $x_1,\dots,x_p$ with maximum value. 
    
    \end{algorithmic}
    \caption{$\frac{1}{H(p)}$-approximation for {\sc $p$-Decomposable-Max-Flow}.}~\label{algo:apx}
    
    \end{algorithm}

    Let ${\cal N}=(D,s,t,c)$ be an instance of the {\sc $p$-Decomposable-Max-Flow} problem. Let $x$ be the flow computed by Algorithm~\ref{algo:apx} on $\mathcal{N}$ and $x^*$ be an optimal solution. Note that $x$ is $p$-decomposable because any flow composed of $i$ arc-disjoint paths in $D_\nu$ is actually a $i$-decomposable flow in $D$.
    
    Assume that $\rho \cdot |x^*| > |x|$ for some $\rho \in [0,1]$, then we will show that $\rho > \frac{1}{H(p)}$ must hold. Observe that, by definition of a $p$-decomposable flow, $x^*$ can be decomposed into $p$ path-flows  $x_1^*,\dots,x_p^*$ of values $c_1^* \geq \dots \geq c_p^*$ respectively (some of them may be empty). For each $i\in [p]$, we define $c_i$ as the largest integer $\nu \in \{1,\dots,c_{\max}\}$ such that $D_\nu$ contains at least $i$ arc-disjoint paths from $s$ to $t$. 

    For each $i\in \{1,\dots,p\}$, let us show that $c_i \geq c_i^*$. If $c_i^* = 0$, this is clear. Else if $c_i^*\geq 1$ then $x_1^*,\dots,x_i^*$ converts into a collection of $i$ non-empty arc-disjoint paths from $s$ to $t$ in $D_{c_i^*}$. Note that they are actually arc-disjoint because each path-flow $x_1^*,\dots,x_i^*$ has value at least $c_i^*$, so if $k$ of them share an arc $uv$, then $c(uv) \geq k \cdot c_i^*$ and $D_{c_i^*}$ contains at least $k$ multiple arcs $uv$. This shows $c_i \geq c_i^*$.
    
    Observe also that $x_i$, the flow computed by Algorithm~\ref{algo:apx} at the $i^{\text{th}}$ step of the first for-loop, has value exactly $i\cdot c_i$. Since $x$ is exactly a flow $x_i$ with maximum value, we deduce the following inequalities for every $i\in [p]$:
    
    \[
        \rho \cdot |x^*| > |x| \geq i\cdot c_i \geq i\cdot c_i^*
    \]
    Multiplying each inequality by $\frac{1}{i}$ and summing all the resulting inequalities, we obtain:
    \[
        \sum_{i=1}^p \left( \frac{1}{i}\cdot \rho \cdot |x^*| \right) > \sum_{i=1}^p c_i^*.
    \]
    Since $|x^*|$ is exactly $\sum_{i} c_i^*$, we deduce from the inequality above that $\rho > \frac{1}{H(p)}$ as desired.
\end{proof}

\begin{theorem}
    \label{thm:apx_disjoint}
    For any fixed $p$, both problems {\sc $p$-Arc-Decomposable-Max-Flow} and {\sc $p$-Vertex-Decomposable-Max-Flow} can be approximated by a ratio $\rho = \frac{1}{H(p)}$.
\end{theorem}
\begin{proof}
    The proofs are really similar the proof of Theorem~\ref{thm:apx_decomposable} so we only briefly describe them.

    For the arc-disjoint version, we will consider Algorithm~\ref{algo:apx} with the following modification. On line~\ref{line:build_Dv}, we build $D_\nu$ by removing every arc $uv$ with capacity $c(uv) < \nu$, and we do not modify the other arcs.
    For the vertex-disjoint version, we do the same modification and we look for vertex-disjoint paths instead of arc-disjoint paths in line~\ref{line:if_arc_disjoint}.

    In both cases, we consider an optimal solution $x^*$ that is the (vertex or arc)-disjoint union of $p$ path-flows $x_1^*,\dots,x_p^*$ (some of them may be empty). The flow sent on each path $x_i^*$ is exactly the minimum capacity of its arc-set. For every $i\in [p]$, if $c_i$ is the largest capacity $\nu$ such that $D_\nu$ contains $i$ (vertex or arc)-disjoint paths, we obtain that $i \cdot c_i \geq i \cdot c_i^*$. We then conclude as in the proof of Theorem~\ref{thm:apx_decomposable}.
\end{proof}

\begin{remark}
    Algorithm~\ref{algo:apx} is a $\frac{2}{3}$-approximation for the {\sc $2$-Decomposable-Max-Flow} problem, and this is best possible by Theorem~\ref{thm:p_decomposable_NPC}.
    For $p=3$ Algorithm 1 is a $\frac{6}{11}$-approximation algorithm and the bound from Theorem~\ref{thm:p_decomposable_NPC} is $\frac{3}{4}$ so there may exist a better approximation algorithm.

\end{remark}
\begin{problem}
    What is the best approximation guarantee one can obtain for the {\sc $p$-Decomposable-Max-Flow} problem when $p>2$?
\end{problem}

\subsection{Restriction to acyclic networks}

In this section, we consider the NP-hard problems of the previous section restricted to acyclic networks. Some remain NP-hard whereas others turn out to be polynomial time solvable.

Given a source $s$ in a digraph $D$ and an ordered set of vertices $W=(v_1,\dots,v_{|W|})$, a \textbf{$W$-tricot} is an ordered set of $|W|$ paths $(Q_1,\dots,Q_{|W|})$ that pairwise intersect exactly on $\{s\}$, and such that $Q_i$ goes from $s$ to $v_i$. The \textbf{value} of a $W$-tricot $T = (Q_1,\dots,Q_{|W|})$ is $(c_1,\dots,c_{|W|})$ where $c_i$ is the minimum capacity along $Q_i$. The \textbf{total value} of $T$ is exactly $\sum_{i=1}^{|W|} c_i$. Given two $W$-tricots $T$ and $T'$ with values $(c_1,\dots,c_{|W|})$ and $(c_1',\dots,c_{|W|}')$ respectively, we consider that the value of $T$ is at least as large as the value of $T'$ if $c_i \geq c_i'$ holds for every $i$.

Given a network $\mathcal{N}=(D,s,t,c)$, if $W\subseteq N^-(t)$, the $(s,t)$-flow associated with a $W$-tricot $T = (Q_1,\dots,Q_{|W|})$ is the flow made by sending exactly $c_i$ units of flow on every path $Q_i$ (extended to $t$) of $T$. Note that this needs to assume that the capacity $c(t_it)$ is at least $c_i$ for each end-vertex $t_i$ of $Q_i$. In Algorithm~\ref{algo:vertex_disjoint_acyclic}, we ensure that this is true by subdividing every arc entering $t$.

\begin{theorem}
    \label{thm:poly-acyclic-disjoint-path}
    When restricted to acyclic networks, the problem {\sc $p$-Vertex-Decomposable-Max-Flow} can be solved in time $O\left(n^{f(p)}\right)$ for some computable function $f$.
\end{theorem}

\begin{proof}
    Our proof is inspired from the one due to Fortune et al.~\cite{fortuneTCS10} when they showed that the \textsc{$p$-Linkage} problem restricted to acyclic networks is in XP when parameterised by $p$.
    Let us show that Algorithm~\ref{algo:vertex_disjoint_acyclic} is correct. Note that this algorithm computes a maximum flow $x$ such that $D_x$ is the vertex-disjoint union of exactly $p$ paths (if such a flow exists). For the general problem, where $D_x$ is the vertex-disjoint union of at most $p$ paths, we just execute Algorithm~\ref{algo:vertex_disjoint_acyclic} for every value $p'\leq p$, and choose the maximum computed flow.

    From now on, we consider $D$ as the digraph where each arc $su$ and $vt$ has been subdivided (after line~1). When we subdivide an arc, we set the capacities of the new arcs to the capacity of the original one. In particular, note that this operation does not change the value of a $p$-vertex-disjoint maximum flow.
    To prove that Algorithm~\ref{algo:vertex_disjoint_acyclic} is correct, we will show by induction on $i\in[r]$ that after iteration $i$ of the for-loop of line~\ref{line:main-for-loop}, every element of $L[W_i]$ is indeed a $W_i$-tricot. Moreover, for every $W_i$-tricot $T$, we will prove that there exists $T'\in L[W_i]$ such that $\text{value}(T') \geq \text{value}(T)$. This will imply the result since, for every $p$-vertex-disjoint flow $x$, its support $D_x$ is actually a $W$-tricot, where $W = V(D_x) \cap N^-(t)$. Let us fix $i\in [r]$ and assume that both statements hold for any $\ell < i$.

    First let $T'$ be any element of $L[W_i]$ that has been added at step $\alpha < i$ of the for-loop of line~\ref{line:main-for-loop}. At iteration $\alpha$, by induction, $T=(Q_1,\ldots{},Q_p)$ is a $W_\alpha$-tricot. Let $v$ be the only vertex in $W_i \setminus W_\alpha$, then $v$ does not belong to any path $Q_h$ otherwise there would be a path from $v$ to the end of $Q_h$, which belongs to $W_\alpha$. This shows that $T'$ must be a $W_i$-tricot.

    Now let $\Tilde{T}$ be any $W_i$-tricot with value $(\tilde{c_1},\dots,\tilde{c_p})$. Let $z$ be the last vertex of $W_i$ according to the computed ordering $v_1,\dots,v_n$. If $z\in N^+(s)$, then $W_i \subseteq N^+(s)$ and the $W_i$-tricot computed in the for-loop of line~\ref{line:for_loop_preprocess} is exactly $\Tilde{T}$.
    Henceforth we assume that $z$ has a predecessor $y \neq s$ in $\Tilde{T}$. Let $W^y$ be the $p$-tuple obtained from $W_i$ by replacing $z$ by $y$. In the computed ordering on the $p$-tuples, $W^y$ must be smaller than $W_i$ (because $z$ is larger than $y$). Let $\ell$ be the index such that $W_\ell = W^y$. By induction, $L[W_\ell]$ contains a $W_\ell$-tricot $T$ with value $(c_1,\dots,c_p)$ at least as large as the value of $\tilde{T} - \{z\}$. Hence, for each $h\in [p]$, $c_h \geq \Tilde{c_h}$. At iteration $\ell$, at some step we consider the tricot $T$, where the vertex $y$ plays the role of $u$ and the vertex $z$ plays the role of $v$. Note that there is no path from $z$ to $W_\ell$, because $z$ is the largest vertex of $W_i$ (according to the computed acyclic ordering) and because $y$ is an in-neighbour of $z$. Hence, at this moment, we consider $W'$ which is actually $W_i$ and $T'$, built from $T$, with value $(c_1',\dots,c_p')$. By construction of $T'$, we have $c_h' = c_h \geq \tilde{c_h}$ when $h\neq j$ and $c_j' = \min \{ c_j, c(yz) \} \geq \Tilde{c_j}$ (where $j$ is the index of the path containing $y$ in $T$, as in the algorithm). Thus, at the end of iteration $\ell$, either $T'$ is an element of $L[W_i]$ larger than $\Tilde{T}$ or $L[W_i]$ already contains an element even larger than $T'$.

    \medskip

    Now we justify that Algorithm~\ref{algo:vertex_disjoint_acyclic} runs in time $O(n^{f(p)})$ for some computable function $f$. Let us bound the number of iterations of each for-loop of the algorithm. Note that the number of $p$-tuples is bounded by $\binom{n+m}{p}$ (recall that we subdivided some arcs in the beginning of the algorithm). Also note that, for every $p$-tuple $W$ and every $W$-tricot $T$ with value $(c_1,\dots,c_p)$, each coordinate $c_i$ must correspond to the capacity of an arc. Thus, the number of possible values for $T$ is bounded by $m^p$. Hence the number of tricots in $L[W]$ is at most $m^p$ since $L[W]$ never contains two $W$-tricots with the same value. 
    Altogether, we get that Algorithm~\ref{algo:vertex_disjoint_acyclic} runs in time $O(n^{f(p)})$ where $f$ is some computable function.
\end{proof}

\begin{algorithm}
\textbf{Input}: A flow network ${\cal N}=(D=(V,A),s,t,c)$ such that $D$ is acyclic.

\textbf{Output}: A maximum flow $x$ such that $D_x$ is the vertex-disjoint union of exactly $p$ paths.

\begin{algorithmic}[1]
\State Subdivide every arc $su$ and $vt$.
\State Tricot$[~]$ $L$ : a list indexed by the $p$-tuples of $V(D)\setminus \{s\}$. Each cell $L[W]$ is a set of $W$-tricots.
\State  Initially, for every $p$-tuple $W$,  $L[W] \gets \emptyset$.
\For{every $p$-tuple $W = (s_1,\dots,s_p) \subseteq N^+(s)$}\label{line:for_loop_preprocess}
    \State $L[W] \gets \{ \text{ the only $W$-tricot $(ss_1,\dots,ss_p)$}  \}$.
\EndFor
\State $v_1,\dots,v_n \gets$ an acyclic ordering of $V(D)$ for which $N^+[s]$ are the first vertices. 
\State $W_1,\dots, W_r \gets$  the lexicographic ordering of the $p$-tuples of $V(D)\setminus \{s\}$ (w.r.t. $v_1,\dots,v_n$).\label{line:ordering_W}
\For{$i=1$ to $r$}\label{line:main-for-loop}
    \For{every tricot $T \in L[W_i]$}
        \State Denote $T=(Q_1,\dots,Q_p)$.
        \For{every vertex $u \in W_i$}\label{line:for-loop-Wi}
            \State $Q_j \gets $ the path of $T$ ending on $u$.
            \For{every vertex $v \in N^+(u)$}\label{line:for-loop-v}
                \If{there is no path from $v$ to $W_i$ in $D$}
                    \State $W' \gets W_i \setminus \{u\} \cup \{v\}$.
                    \State Let $T'=(Q_1',\dots,Q_p')$ be the $W'$-tricot where $Q_j' = Q_j \cup \{v\}$ and $Q_h' = Q_h$ for $h\neq j$.
                    \If{for every $\tilde{T} \in L[W']$ $\text{value}(\Tilde{T}) \not \geq \text{value}(T')$}
                        \State $L[W'] \gets L[W'] \cup T'$.\label{line:change_L_W}
                    \EndIf
                \EndIf
            \EndFor
        \EndFor
    \EndFor
\EndFor
\State Find the tricot $T \in \bigcup_{W\subseteq N^-(t)}L[W]$ with maximum total value.
\State \Return the flow associated with $T$. 

\end{algorithmic}
\caption{{\sc $p$-Vertex-Decomposable-Max-Flow} restricted to acyclic networks.}~\label{algo:vertex_disjoint_acyclic}

\end{algorithm}

\begin{corollary}
    When restricted to acyclic networks, the {\sc $p$-Arc-Decomposable-Max-Flow} problem can be solved in time $O\left(n^{f(p)}\right)$ for some computable function $f$.
\end{corollary}
\begin{proof}
    Let $\mathcal{N}=(D,s,t,c)$ be an instance of the \textsc{$p$-Arc-Decomposable-Max-Flow} problem, with $D$ being acyclic.  We may assume that there is no arc from $s$ to $t$ in $D$, for otherwise we just subdivide it.
    We let $D'$ be the line digraph of $D$, that is $V(D') = A(D)$ and 
    \[
        A(D') = \{ uv \mid u,v\in A(D), \text{ the head of $u$ coincides with the tail of $v$} \}.
    \]
    We then add to $D'$ a source $s'$ and a sink $t'$, and all arcs of
    $\{s'u \mid u \text{ is a leaving arc of $s$ in $D$} \} \cup \{vt' \mid v\in A(D) \text{ is an entering arc of $t$} \}$.
    We finally define the capacities on $A(D')$ as follows:
    \[
        \forall uv\in A(D'), c'(uv) = \left\{
        \begin{array}{ll}
            c(v) & \mbox{if } u=s', \\
            c(u) & \mbox{if } v=t', \\
            \min(c(u),c(v)) & \mbox{otherwise}. \\
        \end{array}
    \right.
    \]
    Let $\mathcal{N}'=(D',s',t',c')$. It is easy to see that we can solve the \textsc{$p$-Arc-Decomposable-Max-Flow} problem on $\mathcal{N}$ by solving the \textsc{$p$-Vertex-Decomposable-Max-Flow} problem on $\mathcal{N}'$. As we saw in Theorem~\ref{thm:poly-acyclic-disjoint-path} this can be done in time $O(|V(D')|^{f(p)})$ so we can solve the \textsc{$p$-Arc-Decomposable-Max-Flow} in time $O(m^{f(p)})=O(n^{f'(p)})$.
\end{proof}

\begin{question}
    Is there an analogue of Theorem~\ref{thm:poly-acyclic-disjoint-path} for {\sc $p$-Decomposable-Max-Flow} ?
\end{question}

\begin{theorem}
    \label{thm:disjoint_acyclic_NPC}
    When $p$ is part of the input, the problems {\sc $p$-Decomposable-Max-Flow}, {\sc $p$-Arc-Decomposable-Max-Flow} and {\sc $p$-Vertex-Decomposable-Max-Flow} are NP-hard for acyclic networks, even when all capacities are 1 or 2.
\end{theorem}
\begin{proof}
We prove the result for the vertex-disjoint version. The hardness of the arc-disjoint version then follows from the usual splitting operation. This also shows the hardness of {\sc $p$-Decomposable-Max-Flow} since, in the reduction, every optimal solution contains exactly one path-flow of value $1$ (and all other path-flows have value 2). Therefore, since all capacities are 1 or 2, the path-flows must be arc-disjoint.

We show that 3-SAT reduces to our problem. The reduction uses a modification of the gadget used in the proof of Theorem 3 in~\cite{evenSJC5}. 
Let ${\cal C}$ be an instance of 3-SAT. We first construct a digraph $\hat{D}_{\cal F}$ which is similar to the digraph $\tilde{D}_{\cal F}$ used in the proof of Theorem~\ref{thm:Delta2acyclicflow}. The only difference is that the digraph $H$ has 5 vertices $y',y,a_1,a_2,a_3$ and the arcs form the three $(y',y)$-paths $y'a_1y,y'a_2y,y'a_3y$.

The analogous version of Remark~\ref{rem:satisfiable} also holds when we want the $(u_1,v_n)$-path in $\tilde{D}_{\cal F}$ to avoid at least one vertex of each set $\{a_{i,1},a_{i,2},a_{i,3}\}$.

Now we construct $\hat{D}_{\cal F}$ from $\tilde{D}_{\cal F}$  by adding two new vertices $s,t$ and the following arcs $su_1,sy'_1,\ldots{},sy'_m, v_nt,y_1t,\ldots{},y_mt$. Finally form the network $\hat{{\cal N}}_{\cal F}=(\hat{D}_{\cal F},s,t,c)$ by giving the arcs $\{sy'_i,y'_ia_{i,k},a_{i,k}y_j,y_jt \mid j\in [m], k\in [3]\}$ capacity 2 and all other arcs capacity 1. Figure~\ref{fig:vertex_disjoint_flow_acyclic} illustrates the construction of $\hat{\mathcal{N}}$.

\ifthenelse{\boolean{colouredfigures}}
{
\begin{figure}[H]
    \begin{minipage}{\linewidth}
        \begin{center}	
          \begin{tikzpicture}[thick,scale=1, every node/.style={transform shape}]
            \tikzset{vertex/.style = {circle,fill=black,minimum size=4pt, inner sep=0pt}}
    	\tikzset{edge/.style = {->,> = latex'}}

            \node[vertex, label=left:$u_1$] (u1) at (2,0) {};
            
            \node[vertex, orange] (xT) at (3.5,1) {};
            \node[vertex,g-green] (xF1) at (3,-1) {};
            \node[vertex,g-blue] (xF2) at (4,-1) {};
            \node[vertex, label=above:$u_2$] (u2) at (5,0) {};
            
            \draw[edge] (u1) -- (xT) node [midway,yshift=0.5em] () {\scriptsize{$1$}};
            \draw[edge] (u1) -- (xF1) node [midway,yshift=-0.5em,xshift=-0.15em] () {\scriptsize{$1$}};
            \draw[edge] (xF1) -- (xF2) node [midway,yshift=-0.5em] () {\scriptsize{$1$}};
            \draw[edge] (xF2) -- (u2) node [midway,yshift=-0.5em,xshift=0.15em] () {\scriptsize{$1$}};
            \draw[edge] (xT) -- (u2) node [midway,yshift=0.5em] () {\scriptsize{$1$}};

            \node[vertex,orange] (yT1) at (6,1) {};
            \node[vertex,g-green] (yT2) at (7,1) {};
            \node[vertex,g-blue] (yF) at (6.8,-1) {};
            \node[vertex, label=above:$u_3$] (u3) at (8,0) {};
            
            \draw[edge] (u2) -- (yF) node [midway,yshift=-0.5em] () {\scriptsize{$1$}};
            \draw[edge] (u2) -- (yT1) node [midway,yshift=0.5em,xshift=-0.15em] () {\scriptsize{$1$}};
            \draw[edge] (yT1) -- (yT2) node [midway,yshift=0.5em] () {\scriptsize{$1$}};
            \draw[edge] (yT2) -- (u3) node [midway,yshift=0.5em,xshift=0.15em] () {\scriptsize{$1$}};
            \draw[edge] (yF) -- (u3) node [midway,yshift=-0.5em] () {\scriptsize{$1$}};

            \node[vertex,g-green] (zT) at (9.5,1) {};
            \node[vertex, orange] (zF1) at (9,-1) {};
            \node[vertex,g-blue] (zF2) at (10,-1) {};
            \node[vertex, label=right:$u_4$] (u4) at (11,0) {};
            
            \draw[edge] (u3) -- (zT) node [midway,yshift=0.5em] () {\scriptsize{$1$}};
            \draw[edge] (u3) -- (zF1) node [midway,yshift=-0.5em,xshift=-0.15em] () {\scriptsize{$1$}};
            \draw[edge] (zF1) -- (zF2) node [midway,yshift=-0.5em] () {\scriptsize{$1$}};
            \draw[edge] (zF2) -- (u4) node [midway,yshift=-0.5em,xshift=0.15em] () {\scriptsize{$1$}};
            \draw[edge] (zT) -- (u4) node [midway,yshift=0.5em] () {\scriptsize{$1$}};

            \node[vertex, label=left:$y_1$, orange] (y1) at (4.25,-2.5) {};
            \node[vertex, label=left:$y_2$, g-green] (y2) at (6.5,-2.5) {};
            \node[vertex, label=left:$y_3$, g-blue] (y3) at (8.75,-2.5) {};
            \node[vertex, label=below:$t$] (t) at (6.5,-4.5) {};
            \draw[edge, orange] (xT) -- (y1);
            \draw[edge, orange] (yT1) -- (y1);
            \draw[edge, orange] (zF1) -- (y1);
            \draw[edge, orange] (y1) -- (t) node [midway,xshift=-1em] () {\scriptsize{$2$}};
            \draw[edge, g-green] (xF1) -- (y2);
            \draw[edge, g-green] (yT2) -- (y2);
            \draw[edge, g-green] (zT) -- (y2);
            \draw[edge, g-green] (y2) -- (t) node [midway,xshift=-0.6em] () {\scriptsize{$2$}};
            \draw[edge, g-blue] (xF2) -- (y3);
            \draw[edge, g-blue] (yF) -- (y3);
            \draw[edge, g-blue] (zF2) -- (y3);
            \draw[edge, g-blue] (y3) -- (t) node [midway,xshift=1em] () {\scriptsize{$2$}};
            \draw[edge, purple] (u4) to[in=0, out=-90] (t);
            \node[purple] (tcapacity) at (9.7,-3.5) {\scriptsize{$1$}};

            \node[vertex, label=left:$y_1'$, orange] (y1s) at (4.25,2.5) {};
            \node[vertex, label=left:$y_2'$, g-green] (y2s) at (6.5,2.5) {};
            \node[vertex, label=left:$y_3'$, g-blue] (y3s) at (8.75,2.5) {};
            \node[vertex, label=above:$s$] (s) at (6.5,4.5) {};
            \draw[edge, orange] (y1s) -- (xT);
            \draw[edge, orange] (y1s) -- (yT1);
            \draw[edge, orange] (y1s) -- (zF1);
            \draw[edge, orange] (s) -- (y1s) node [midway,xshift=-1em] () {\scriptsize{$2$}};
            \draw[edge, g-green] (y2s) -- (xF1);
            \draw[edge, g-green] (y2s) -- (yT2);
            \draw[edge, g-green] (y2s) -- (zT);
            \draw[edge, g-green] (s) -- (y2s) node [midway,xshift=-0.6em] () {\scriptsize{$2$}};
            \draw[edge, g-blue] (y3s) -- (xF2);
            \draw[edge, g-blue] (y3s) -- (yF);
            \draw[edge, g-blue] (y3s) -- (zF2);
            \draw[edge, g-blue] (s) -- (y3s) node [midway,xshift=1em] () {\scriptsize{$2$}};
            \draw[edge, purple] (s) to[in=90, out=180] (u1);
            \node[purple] (scapacity) at (2.7,3.5) {\scriptsize{$1$}};
          \end{tikzpicture}
      \caption{The network $\hat{\mathcal{N}}_\mathcal{F}$ when $\mathcal{F} = \textcolor{orange}{(x_1 \vee x_2 \vee \neg x_3)} \wedge \textcolor{g-green}{(\neg x_1 \vee x_2 \vee x_3)} \wedge \textcolor{g-blue}{(\neg x_1 \vee \neg x_2 \vee \neg x_3)}$.}
      \label{fig:vertex_disjoint_flow_acyclic}
    \end{center}    
  \end{minipage}
\end{figure}
}
{
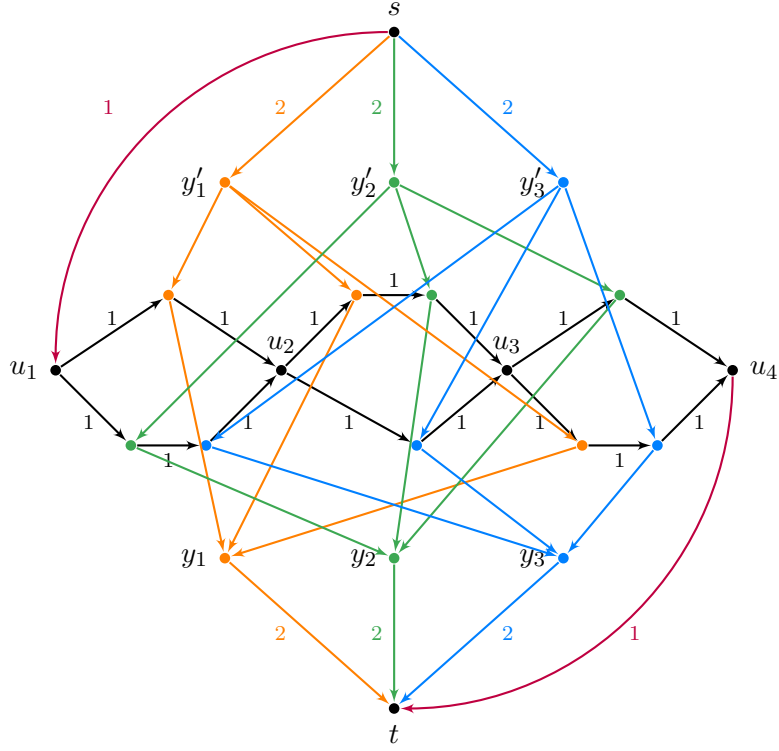
\begin{figure}
    \begin{minipage}{\linewidth}
        \begin{center}	
          \begin{tikzpicture}[thick,scale=1, every node/.style={transform shape}]
            \tikzset{vertex/.style = {circle,fill=black,minimum size=4pt, inner sep=0pt}}
    	\tikzset{edge/.style = {->,> = latex'}}

            \node[vertex, label=left:$u_1$] (u1) at (2,0) {};
            
            \node[vertex] (xT) at (3.5,1) {};
            \node[vertex] (xF1) at (3,-1) {};
            \node[vertex] (xF2) at (4,-1) {};
            \node[vertex, label=above:$u_2$] (u2) at (5,0) {};
            
            \draw[edge] (u1) -- (xT) node [midway,yshift=0.5em] () {\scriptsize{$1$}};
            \draw[edge] (u1) -- (xF1) node [midway,yshift=-0.5em,xshift=-0.15em] () {\scriptsize{$1$}};
            \draw[edge] (xF1) -- (xF2) node [midway,yshift=-0.5em] () {\scriptsize{$1$}};
            \draw[edge] (xF2) -- (u2) node [midway,yshift=-0.5em,xshift=0.15em] () {\scriptsize{$1$}};
            \draw[edge] (xT) -- (u2) node [midway,yshift=0.5em] () {\scriptsize{$1$}};

            \node[vertex] (yT1) at (6,1) {};
            \node[vertex] (yT2) at (7,1) {};
            \node[vertex] (yF) at (6.8,-1) {};
            \node[vertex, label=above:$u_3$] (u3) at (8,0) {};
            
            \draw[edge] (u2) -- (yF) node [midway,yshift=-0.5em] () {\scriptsize{$1$}};
            \draw[edge] (u2) -- (yT1) node [midway,yshift=0.5em,xshift=-0.15em] () {\scriptsize{$1$}};
            \draw[edge] (yT1) -- (yT2) node [midway,yshift=0.5em] () {\scriptsize{$1$}};
            \draw[edge] (yT2) -- (u3) node [midway,yshift=0.5em,xshift=0.15em] () {\scriptsize{$1$}};
            \draw[edge] (yF) -- (u3) node [midway,yshift=-0.5em] () {\scriptsize{$1$}};

            \node[vertex] (zT) at (9.5,1) {};
            \node[vertex] (zF1) at (9,-1) {};
            \node[vertex] (zF2) at (10,-1) {};
            \node[vertex, label=right:$u_4$] (u4) at (11,0) {};
            
            \draw[edge] (u3) -- (zT) node [midway,yshift=0.5em] () {\scriptsize{$1$}};
            \draw[edge] (u3) -- (zF1) node [midway,yshift=-0.5em,xshift=-0.15em] () {\scriptsize{$1$}};
            \draw[edge] (zF1) -- (zF2) node [midway,yshift=-0.5em] () {\scriptsize{$1$}};
            \draw[edge] (zF2) -- (u4) node [midway,yshift=-0.5em,xshift=0.15em] () {\scriptsize{$1$}};
            \draw[edge] (zT) -- (u4) node [midway,yshift=0.5em] () {\scriptsize{$1$}};

            \node[vertex, label=left:$y_1$] (y1) at (4.25,-2.5) {};
            \node[vertex, label=left:$y_2$] (y2) at (6.5,-2.5) {};
            \node[vertex, label=left:$y_3$] (y3) at (8.75,-2.5) {};
            \node[vertex, label=below:$t$] (t) at (6.5,-4.5) {};
            \draw[edge,  densely dotted] (xT) -- (y1);
            \draw[edge,  densely dotted] (yT1) -- (y1);
            \draw[edge,  densely dotted] (zF1) -- (y1);
            \draw[edge,  densely dotted] (y1) -- (t) node [midway,xshift=-1em] () {\scriptsize{$2$}};
            \draw[edge, loosely dotted] (xF1) -- (y2);
            \draw[edge, loosely dotted] (yT2) -- (y2);
            \draw[edge, loosely dotted] (zT) -- (y2);
            \draw[edge, loosely dotted] (y2) -- (t) node [midway,xshift=-0.6em] () {\scriptsize{$2$}};
            \draw[edge, dashed] (xF2) -- (y3);
            \draw[edge, dashed] (yF) -- (y3);
            \draw[edge, dashed] (zF2) -- (y3);
            \draw[edge, dashed] (y3) -- (t) node [midway,xshift=1em] () {\scriptsize{$2$}};
            \draw[edge ] (u4) to[in=0, out=-90] (t);
            \node[] (tcapacity) at (9.7,-3.5) {\scriptsize{$1$}};

            \node[vertex, label=left:$y_1'$] (y1s) at (4.25,2.5) {};
            \node[vertex, label=left:$y_2'$] (y2s) at (6.5,2.5) {};
            \node[vertex, label=left:$y_3'$] (y3s) at (8.75,2.5) {};
            \node[vertex, label=above:$s$] (s) at (6.5,4.5) {};
            \draw[edge,  densely dotted] (y1s) -- (xT);
            \draw[edge,  densely dotted] (y1s) -- (yT1);
            \draw[edge,  densely dotted] (y1s) -- (zF1);
            \draw[edge,  densely dotted] (s) -- (y1s) node [midway,xshift=-1em] () {\scriptsize{$2$}};
            \draw[edge, loosely dotted] (y2s) -- (xF1);
            \draw[edge, loosely dotted] (y2s) -- (yT2);
            \draw[edge, loosely dotted] (y2s) -- (zT);
            \draw[edge, loosely dotted] (s) -- (y2s) node [midway,xshift=-0.6em] () {\scriptsize{$2$}};
            \draw[edge, dashed] (y3s) -- (xF2);
            \draw[edge, dashed] (y3s) -- (yF);
            \draw[edge, dashed] (y3s) -- (zF2);
            \draw[edge, dashed] (s) -- (y3s) node [midway,xshift=1em] () {\scriptsize{$2$}};
            \draw[edge ] (s) to[in=90, out=180] (u1);
            \node[] (scapacity) at (2.7,3.5) {\scriptsize{$1$}};
          \end{tikzpicture}
      \caption{The network $\hat{\mathcal{N}}_\mathcal{F}$ when $\mathcal{F} = {(x_1 \vee x_2 \vee \neg x_3)} \wedge (\neg x_1 \vee x_2 \vee x_3) \wedge (\neg x_1 \vee \neg x_2 \vee \neg x_3)$.}
      \label{fig:vertex_disjoint_flow_acyclic}
    \end{center}    
  \end{minipage}
\end{figure}
}

We claim that $\hat{{\cal N}}_{\cal F}$ has an $(s,t)$-flow of value $2m+1$ which can be decomposed into $m+1$ vertex-disjoint path-flows if and only if ${\cal F}$ is satisfiable.

Suppose first that $\phi$ is a truth assignment which satisfies ${\cal F}$. Fix one true literal for each clause and let $\ell_i\in [3]$ be the index of the true literal we chose for $C_i$. Let $P$ be the $(u_1,v_n)$-path which for each $j\in [n]$ follows the subpath $u_jz_1\ldots{}z_{q_j}v_j$ if $\phi(x_j)$ is true and otherwise follows the subpath $u_jy_1\ldots{}y_{p_j}v_j$. Now we can send 2 units of flow along each of the paths $sy'_ia_{i,\ell_i}y_it$ and one unit along the path $sPt$. By construction, all the $m+1$ paths we used are vertex-disjoint.

Suppose now that $\hat{{\cal N}}_{\cal F}$ has an $(s,t)$-flow of value $2m+1$ which can be decomposed into $m+1$ vertex-disjoint path-flows. Let $sQt$ be the path on which we send one unit of flow and note that $Q$ must be a $(u_1,v_n)$-path. For each $j\in [n]$ this  path will use either the subpath $u_jy_1\ldots{}y_{p_j}v_j$ or the subpath $u_jz_1\ldots{}z_{q_j}v_j$. If $Q$ uses the first subpath we set $x_j$ to false and otherwise we set $x_j$ to true. Since we have $m$ internally disjoint $(s,t)$-paths which avoid all vertices of $Q$, for each $i\in [m]$, at least one of the $a_{i,h}$s where $h\in [3]$ is not on $Q$. By the analogous version of Remark~\ref{rem:satisfiable}, this shows that our truth assignment satisfies ${\cal F}$.
\end{proof}

The following result shows that Theorem~\ref{thm:poly-acyclic-disjoint-path} is somehow best possible, in the sense that \textsc{$p$-Vertex-Decomposable-Max-Flow}, restricted to acyclic digraphs, cannot be solved in time $O\left(f(p)\cdot n^{O(1)}\right)$ for any computable function $f$, unless FPT = W[1].

\begin{theorem}
    The three problems
    \begin{itemize}
        \item {\sc $p$-Vertex-Decomposable-Max-Flow},
        \item {\sc $p$-Arc-Decomposable-Max-Flow}, and
        \item {\sc $p$-Decomposable-Max-Flow},
    \end{itemize}
    are all W[1]-hard when parameterised by $p$.
\end{theorem}
\begin{proof}
    We first show that the \textsc{$p$-Vertex-Decomposable-Max-Flow} problem on acyclic digraphs is W[1]-hard when parameterised by $p$ by reducing from \textsc{$p$-Linkage} parameterised by $p$ on acyclic digraphs. This problem is W[1]-hard by Theorem~\ref{thm:k_linkage_W1_hard}. We will then show how to adapt it for \textsc{$p$-Arc-Decomposable-Max-Flow} and \textsc{$p$-Decomposable-Max-Flow}.

    Let $\mathcal{I}=(D,(s_i)_{i\in[p]},(t_i)_{i\in[p]})$ be an instance of the {\sc $p$-Linkage} problem for acyclic digraphs. We first add a source $s$ and a sink $t$ to $D$, and all the arcs $ss_i,t_it$ for $i\in [p]$. Clearly, the obtained digraph remains acyclic. For every $i\in [p]$, the capacity of $ss_i$ and $t_it$ is $i$. We set the capacity of every other arc to $p$.

    Let $\mathcal{N}$ be the obtained network. It is clear that $\mathcal{I}$ is a positive instance if and only if the solution of \textsc{$p$-Vertex-Decomposable-Max-Flow} on $\mathcal{N}$ is $\frac{1}{2}p(p+1)$. This shows the result.

    For \textsc{$p$-Arc-Decomposable-Max-Flow}, we do exactly the same reduction but we reduce from {\sc Weak-$p$-Linkage}. For \textsc{$p$-Decomposable-Max-Flow}, we reduce from {\sc Weak-$p$-Linkage} but we change the capacities to ensure that the path-flows are pairwise disjoint. So we set the capacities of $ss_i$ and $t_it$ to $p+i$ for every $i\in [p]$, and we set the capacity of every other arc to $2p$.
\end{proof}

\section{Separable flows}\label{sec:sepflow}

Let $x$ be a flow in a network $\mathcal{N}=(D,s,t,c)$ that is the union of $p$ path-flows $Q_1,\dots,Q_p$. Then $x$ is a \textbf{$q$-vertex-separable} flow if each vertex of $V(D)\setminus \{s,t\}$ belongs to at most $q$ different paths $Q_i$. Analogously, $x$ is a \textbf{$q$-arc-separable} flow if each arc of $D$ belongs to at most $q$ different paths $Q_i$. We consider the following two problems.

\defproblem
{\sc $q$-Vertex-Separable-Max-Flow}
{A flow network ${\cal N}=(D,s,t,c)$.}
{The maximum value of a flow $x$ such that $x$ can be decomposed into path-flows with the property that each vertex of $D-\{s,t\}$ belongs to at most $q$ such path-flows.}

\defproblem
{\sc $q$-Arc-Separable-Max-Flow}
{A flow network ${\cal N}=(D,s,t,c)$.}
{The maximum value of a flow $x$ such that $x$ can be decomposed into path-flows with the property that each arc of $D$ belongs to at most $q$ such path-flows.}

Note that a flow $x$ is $1$-vertex separable if it is made of $p$ path-flows $Q_1,\dots,Q_p$ that are pairwise intersecting exactly on $\{s,t\}$. Analogously, $x$ is $1$-arc separable if $Q_1,\dots,Q_p$ are arc-disjoint. Note that the number $p$ of paths is not constrained. 
In the following we show that the two problems above are NP-hard even when the network is acyclic and capacities are in $\{1,2\}$.
This is in contrast with Theorem~\ref{thm:poly-acyclic-disjoint-path}: unless P=NP, neither {\sc $q$-Arc-Separable-Max-Flow} nor {\sc $q$-Vertex-Separable-Max-Flow}, restricted to acyclic networks, can be solved in time $O(n^{f(q)})$ for any computable function $f$.

\begin{theorem}
    For every fixed $q\geq 1$, both {\sc $q$-Arc-Separable-Max-Flow} and {\sc $q$-Vertex-Separable-Max-Flow} are NP-hard even when restricted to acyclic networks with capacities $\{1,2\}$.
\end{theorem}
\begin{proof}
    We will show the result for {\sc $q$-Vertex-Separable-Max-Flow}. The hardness of {\sc $q$-Arc-Separable-Max-Flow} then follows easily by the usual splitting operation.

    Let us fix $q$. We will reduce from {\sc $p$-Vertex-Decomposable-Max-Flow}, with $p$ being part of the input, which is NP-hard on acyclic networks with capacities in $\{1,2\}$ by Theorem~\ref{thm:disjoint_acyclic_NPC}. The construction in the proof of Theorem~\ref{thm:disjoint_acyclic_NPC} shows that the problem remains hard if $p$ is equal to the out-degree of $s$. 

    So let $\mathcal{N}=(D,s,t,c)$ be a network instance of {\sc $p$-Vertex-Decomposable-Max-Flow} with $p$ being equal to the out-degree of $s$ in $D$,  $D$ is acyclic and the capacities are in $\{1,2\}$. We build $D'$ from $D$ as follows. For every vertex $v\in V(D)\setminus \{s,t\}$, we add to $D$ $2(q-1)$ new vertices $v_1^-,\dots,v_{q-1}^-,v_1^+,\dots,v_{q-1}^+$. Then we add every arc of the path $sv_i^-vv_{i}^+t$ (for $i\in [q-1]$). Note that $D'$ remains acyclic. Then we form $\mathcal{N}'=(D',s,t,c)$ by giving all the new arcs capacity 2.

    We claim that $\mathcal{N}$ admits an ($s,t$)-flow $x$ made of at most $p$ internally vertex-disjoint paths of value $|x|$ if and only if $\mathcal{N}'$ admits a $q$-vertex-separable ($s,t$)-flow $x'$ of value $|x'| = |x| + 2n(q-1)$, where $n= |V(D) \setminus \{s,t\}|$. This equivalence shows the result.

    \medskip

    Assume first that $\mathcal{N}$ admits an ($s,t$)-flow $x$ made of at most $p$ internally vertex-disjoint path-flows. We call these path-flows as the original ones. Note that this flow is also a flow of $\mathcal{N}'$. We complete this flow by sending 2 units of flow along every path of the form $sv_i^-vv_{i}^+t$ (for $v\in V(D)\setminus \{s,t\}$ and $i\in [q-1]$). We call these $n(q-1)$ path-flows the new ones. The obtained flow $x'$ has value $|x|+2n(q-1)$ as desired. Since the original path-flows are disjoint, and because every vertex (different from $s,t$) belongs to at most $q-1$ new path-flows, we get that every vertex belongs to at most $q$ path-flows. Hence $x'$ is $q$-vertex-separable.

    \medskip

    Conversely, assume now that $\mathcal{N}'$ admits a $q$-separable flow of value $\nu+2n(q-1)$. Since it is $q$-separable, it is decomposable into path-flows such that each vertex but $s,t$ belongs to at most $q$ of them. Let $x'$ be such a flow and $Q_1',\dots,Q_{p'}'$ be such a decomposition into path-flows for which the number of path-flows of value 2 of the form $sv^-_ivv^+_it$ is maximized.
    It is straightforward that, in this case, the number of such path-flows is exactly $n(q-1)$. Let $Q_1,\dots,Q_{p'-n(q-1)}$ be the other path-flows, then $p'-n(q-1) \leq p$ because $p$ is the out-degree of $s$ in $D$. We claim that there are pairwise disjoint (except on $\{s,t\}$). Assume not, meaning that $Q_i$ and $Q_j$ are intersecting on $v \in V(D) \setminus \{s,t\}$. Note that $v$ belongs to $(q-1)$ other path-flows of $Q_1',\dots,Q_{p'}'$ (the ones of the form $sv^-_ivv^+_it$). Thus $v$ belongs to $q+1$ path-flows of $Q_1',\dots,Q_{p'}'$, a contradiction. Finally note that the flow made of the path-flows $Q_1,\dots,Q_{p'-n(q-1)}$ has value $\nu$, which concludes the proof.
\end{proof}

\section*{Acknowledgments}
The authors are thankful to the anonymous referees who went deeply into the proofs of the paper, and for their valuable comments, which improved the quality of the manuscript.

\bibliography{refs}

\end{document}